\documentclass{JINST}

\def\rp{\hbox{$\cal P $}} 
\def\rw{\hbox{$\cal W $}} 

\preprint{LAL 10-43}

\title{A High Precision Fabry-Perot Cavity Polarimeter at HERA}

\author{S.~Baudrand, M.~Bouchel, V.~Brisson, R. Chiche, M.~Jacquet, 
 S.~Kurbasov\thanks{On leave of absence from P.N.~Lebedev Physical Institute, 
 Moscow, Russian Federation.} ,
 G.~Li\thanks{Now at Institute of High Energy Physics, Chinese Academy of 
 Science, Beijing, China.}, C.~Pascaud, A.~Reboux, V.~Soskov, 
 Z.~Zhang\thanks{Corresponding author.}, F.~Zomer\\
\llap{}Laboratoire de l'Acc\'el\'erateur Lin\'eaire, Univ.\ Paris-Sud et 
IN2P3/CNRS,\\
  Orsay, France\\
  E-mail: \email{zhangzq@lal.in2p3.fr}}

\author{M.~Beckingham\thanks{Now at Albert-Ludwigs-University Freiburg, Freiburg, Germany.}, T.~Behnke, N.~Coppola\thanks{Now at European XFEL GmbH, Hamburg, Germany.}, N.~Meyners, D.~Pitzl, 
 S.~Schmitt\\
\llap{}DESY\\
  Hamburg, Germany}

\author{M. Authier, P. Deck-Betinelli\thanks{Now at Synchrotron SOLEIL, 
 Saint Aubin, France.}, Y. Queinec\\
\llap{}CEA, IRFU, SIS, Centre de Saclay\\
  Gif sur Yvette, France}

\author{L. Pinard\\
\llap{}LMA, Universit\'e Claude Bernard LYON I et IN2P3/CNRS\\
  Villeurbanne, France.}

\abstract{A Fabry-Perot cavity polarimeter, installed in 2003 at HERA for  
the second phase of its operation, is described.
The cavity polarimeter was designed to measure the longitudinal polarisation
of the HERA electron beam
with high precision for each electron bunch spaced with a time interval  
of $96\,{\rm ns}$.
Within the cavity the laser intensity was routinely enhanced 
up to a few ${\rm kW}$ from its original value of $0.7\,{\rm W}$  
in a stable and controllable way. 
By interacting such a high intensity laser beam with the HERA electron beam 
it is possible to measure its polarisation with a relative statistical 
precision of $2\%$ per bunch per minute. 
Detailed systematic studies have also been performed resulting in  
a systematic uncertainty of $1\%$.}

\keywords{Polarimeter, Fabry-Perot cavity}

\begin{document}

\section{Introduction} 
The electron\footnote{The term ``electron'' is used generically to refer to 
both electrons and positrons, unless otherwise stated.}-proton collider HERA  
was upgraded after  
a first phase of data taking in years $1992-2000$. The aim of the upgrade
was to increase  
the luminosity by a factor of three and to provide a longitudinally  
polarised electron beam to the two general purpose detector experiments  
H1~\cite{h1} and ZEUS~\cite{zeus}, in addition to the fixed target experiment  
HERMES~\cite{hermes}. 

To cope with the physics program after the upgrade, a fast and high precision 
longitudinal Compton polarimeter using a continuous wave laser resonating in a
Fabry-Perot cavity (LPOL cavity) was proposed~\cite{kogelnik},
constructed and installed  
near the existing longitudinal Compton polarimeter (LPOL)~\cite{lpol}. 
In addition to the LPOL, the transverse polarisation of the electron beam is 
also measured by another polarimeter (TPOL)~\cite{barber93}. 

With respect to the prior HERA LPOL and TPOL polarimeters, the
higher statistical precision of the LPOL cavity is achieved by  
increasing firstly the power of the continuous wave laser by more than 
two orders of magnitude compared to the TPOL and  
secondly the frequency of the electron-photon(laser) interaction  
to $10\,{\rm MHz}$ compared to $0.1\,{\rm kHZ}$ of the pulsed laser of the 
LPOL. 
A new Data Acquisition System (DAQ), synchronised to the HERA beam clock, 
has been developed accordingly which operates without any trigger at 
$10\,{\rm MHz}$.
This is one of the novelties of the experiment described in this article. 

The HERA Fabry-Perot cavity is similar to a device that 
has been used successfully to measure 
the polarisation of the CEBAF LINAC electron beam~\cite{nicoet,bardin,bardin2}. 
One major difference between the HERA and CEBAF LPOL cavities is 
the dynamical regime.
Whereas the luminosity of Compton scattering is relatively low at CEBAF, 
it reaches much higher values at HERA. That is, the average number of 
scattered Compton photons is close to one per bunch in the latter case and 
much smaller in the former.  
The HERA dynamical regime, denoted as `few photon mode' in this article, 
has been used successfully for the first time by the LPOL cavity
to measure the electron beam polarisation.
An important point to mention is that a huge effort was made to reduce 
the unforeseen high level of synchrotron radiation emitted by the electron
beam at the cavity location downstream of the HERMES experiment, 
by adding many protections (cf Sect.~\ref{sec:design} for more detail) 
to avoid damaging any optical or electronic component~\cite{sb07}. 
Especially fragile were the `supermirrors',
whose high quality, which was maintained  
until the end of data taking, was the key to obtaining the foreseen 
high power of the laser,

The main purpose of this article is to describe the LPOL cavity experiment and
to report about the electron polarisation measurement.
The article is organised as follows: In Sect.~\ref{sec:pola-hera},  
the principle of the electron beam polarisation measurement at HERA 
is discussed; In Sect.~\ref{sec:cavity}, the experimental setup of  
the LPOL cavity is described; Sections~\ref{sec:method}-\ref{sec:syst} are
dedicated to the analysis method, the data taking, 
the polarisation results and the systematic studies, respectively,
followed by Sect.~\ref{sec:summary} for summary.

\section{Polarisation at HERA}\label{sec:pola-hera} 
Both the electron and proton ring of the HERA collider can have up to 
$220$ bunches. Most of the electron and proton bunches are filled to 
allow the $ep$ collisions. These are so-called colliding bunches. 
However, a small fraction of proton bunches 
are unfilled and thus the corresponding electron bunches have no partner.
These are non-colliding or pilot electron bunches.
The beam polarisation from both bunch types have to be measured.

In a storage ring like HERA, the electron beam 
has a natural transverse polarisation caused 
by emission of synchrotron radiation due to  
the bending magnetic field. This is the so-called Sokolov-Ternov (ST)  
effect~\cite{LKST}. 
The transverse polarisation arises from a small difference in the spin flip  
probabilities during a complete turn between the up-to-down 
$(w_{\uparrow\downarrow}$) and down-to-up ($w_{\downarrow\uparrow}$) flips. 
The evolution of the spin up and spin down electron populations is given
by
\begin{equation} 
\frac{\partial \rp_\uparrow}{\partial t} = -
\frac{\partial \rp_\downarrow}{\partial t} = 
w_{\downarrow\uparrow} \rp_\downarrow
-w_{\uparrow\downarrow} \rp_\uparrow\,.
\end{equation}
Solving this differential equation leads to the polarisation evolution
\begin{equation} 
P(t)=P^\infty \left(1-e^{-\frac{t}{\tau}}\right) 
\end{equation}
where
\begin{equation}
P(t)=\frac{\rp_\uparrow-\rp_\downarrow}{\rp_\uparrow+\rp_\downarrow}\,,
\hspace{15mm}
P^\infty=\frac{w_{\uparrow\downarrow}-w_{\downarrow\uparrow}} 
            {w_{\uparrow\downarrow}+w_{\downarrow\uparrow}}\,,
\hspace{15mm}
\tau=\frac{C} {w_{\uparrow\downarrow}+w_{\downarrow\uparrow}}
\end{equation}
with $P^\infty$ and $\tau$ being the maximum polarisation value and
the intrinsic rise-time, respectively, and $C$ is a constant depending on 
the ring parameters.

For a perfect flat HERA machine, where the polarisation of the beam is only 
due to the ST effect, one has asymptotically\footnote{Radiation 
depolarisation may result from quantum phase jumps~\cite{barber99}.}
$P^\infty=P_{\rm ST}=\frac{8}{5\sqrt{3}}=92.4\%$ 
and $\tau=\tau_{\rm ST}\simeq 36.5\,{\rm min}$~\cite{barber96}.
However as most of the physics at HERA is sensitive to 
longitudinal polarisation one needs to insert 
a device that transforms the transverse polarisation into longitudinal. 
This kind of device is called spin rotator and in the case of 
HERA a ``Mini-Rotator'' design, developed by Buon and Steffen~\cite{spinrot}, 
has been used.  
The spin rotators are located around the electron-proton interaction points 
and are installed in pairs allowing transversely polarised positrons 
to be rotated into longitudinally polarised states and back again.

These spin rotators are however responsible for a depolarisation 
which increases $w_{\uparrow\downarrow}$ and $w_{\downarrow\uparrow}$.
Nevertheless as the effect is mainly geometrical one expects that
the difference $w_{\uparrow\downarrow}-w_{\downarrow\uparrow}$ is
largely preserved, meaning that
$P^\infty/\tau$ stays constant.
The validity of this assumption may be tested by measuring the rise time and
maximum polarisation in a rise-time experiment for both a flat and
standard ring. For HERA it is expected that
\begin{equation} 
\frac{P^\infty}{\tau}=\frac{P_{ST}}{\tau_{ST}}=0.02532\,({\min}^{-1})\,.
\label{eq:st}
\end{equation}

\subsection{Polarisation Measurement - Polarimeters} 
Compton-laser polarimeters, widely used at $\rm{e}^+\rm{e}^-$ 
storage rings (e.g.\ LEP~\cite{lep-tpol}, TRISTAN~\cite{tristan}), 
are based on the spin  
dependent cross section for Compton scattering of  
polarised photons on electrons.
The longitudinal component of  
the electron polarisation is measured through the energy  
dependence of the cross section. The polarisation  
can be deduced from measurements of the final state 
electron~\cite{sld,slacupgrade},
but at storage rings it is more practical to  
detect the scattered photon.  
In this article we shall concentrate on the measurement of longitudinal 
polarisation of the electron beam.

Assuming a mono-energetic and mono-directional electron beam interacting 
with a laser beam, the number of scattered photons per unit of 
time and solid angle in the electron rest frame (with the $z$ axis 
in the direction of motion of the electron beam) is given 
by~\cite{fano,barber93} 
\begin{eqnarray} 
\frac{d^3n_\gamma}{dtd\Omega}={\cal L}_{e\gamma}C\biggl\{&&\hspace{-5mm}
\left[1+\cos^2\theta+2(k_i-k_f)\sin^2\frac{\theta}{2}\right]\nonumber\\ 
&-&[S_1\cos2\phi+S_2 \sin2\phi ]\sin^2\theta\nonumber\\ 
&-&2\sin\theta\sin^2\frac{\theta}{2}S_3[P_y\sin\phi-P_x\cos\phi]\nonumber\\ 
&-&2\cos\theta\sin^2\frac{\theta}{2}(k_f+k_i)S_3P_z 
\biggr\}\,,\label{Compton-xsec} 
\end{eqnarray}  
with~\cite{refjapon}
\begin{equation}\label{eglumi} 
{\cal L}_{e\gamma}\approx \frac{1}{\sqrt{2\pi}} 
                        \frac{1+\cos\alpha_{e\gamma}}{\sin\alpha_{e\gamma}} 
                        \frac{I_e}{ec} 
                        \frac{P_{\rm laser}\lambda}{hc} 
                        \frac{1}{\sigma^2_{e,x}+\sigma^2_\gamma} 
\hspace{5mm} 
C=\frac{1}{2}\left(\frac{e^2}{m_ec^2}\frac{k_f}{k_i}\right)^2\,, 
\end{equation} 
where in Eq.(\ref{Compton-xsec}), $k_i$ and $k_f$ are the momenta 
of the incident and scattered photons in the electron rest frame,  
$\theta$ is the angle between the two and $\phi$ the azimuthal angle 
in ($x$,$y$) plane perpendicular to the electron beam line axis $z$;
$\cos\theta$ is univoquely related to the energy of
the photon in the laboratory frame so that Eq.(\ref{Compton-xsec}) 
also represents 
the scattered photon energy spectrum. The
variables $P_x$, $P_y$ and $P_z$ denote the three components of 
the electron polarisation vector in the $x$, $y$ and $z$ directions. 
The circular photon polarisation is described by the third component
$S_3$ of the Stokes parameters ($S_1$, $S_2$, $S_3$)~\cite{fano}.
The electron beam-laser beam luminosity ${\cal L}_{e\gamma}$ is given 
in Eq.(\ref{eglumi}) and the other parameters in Eq.(\ref{eglumi}) are 
the electron-laser crossing angle $\alpha_{e\gamma}$,
the electron beam current $I_e$, transverse beam size $\sigma_{e,x}$ in $x$, 
the laser beam power $P_{\rm laser}$, wavelength $\lambda$, beam size 
$\sigma_{\gamma}$, the electron electric charge $e$, mass $m_e$ and 
the speed of light $c$. 

Since the nominal HERA electron beam energy is around $27.5\,{\rm GeV}$, 
much larger than that of the laser beam, the photons are scattered within 
a tiny cone of a few hundreds of micro-radian
in the direction of the electron 
beam. Therefore the photon energy distribution can be measured  
within a small calorimeter. 

The three components of the electron polarisation  
appear only in the third and fourth lines on the right-hand 
side of Eq.(\ref{Compton-xsec}) 
and are connected with the circular laser polarisation component 
$S_3$ only.  
Therefore for a precise determination of the electron polarisation, one 
needs to maximize the level of circular laser polarisation  
($S_3 \rightarrow \pm 1$). Knowing $S_3$, 
the electron longitudinal polarisation $P_z$ can be determined by a fit  
to the distribution of the scattered photon energy.  
To determine the transverse polarisation, one has to measure both 
the energy distribution and the azimuthal angle $\phi$.

\subsection{Polarisation Measurement Modes} \label{sec:meas_mode}

At HERA, where the electron bunches are separated by $\Delta t=96\,{\rm ns}$ 
in time, the number of back-scattered Compton photons (hereafter named BCP) 
per bunch is given by 
$n_\gamma=\Delta t\int d\Omega (d^3n_\gamma/dtd\Omega)$.  
Depending on the value of $n_\gamma$, one can define three different 
measurement modes: (1) single photon mode ($n_\gamma \ll 1$),  
(2) few photon mode ($n_\gamma\approx 1$) and  
(3) multi-photon mode ($n_\gamma \gg 1$). 
These are the operation modes of the TPOL~\cite{barber93}, 
the LPOL cavity and the existing LPOL~\cite{lpol} respectively. 

The advantage of the single photon mode is that one can calibrate  
the calorimeter in an absolute way using two reference points of the photon 
energy spectrum independently of the electron beam polarisation:
firstly the Compton kinematic edge, located at
$10\,{\rm GeV}$ at HERA for a $1\,064\,{\rm nm}$ laser beam wavelength, 
and secondly the bremsstrahlung kinematic edge, located  
around $27.5\,{\rm GeV}$ which corresponds to photons radiated from 
the scattering  of the electron 
beam with the residual gas of the vacuum beam pipe (hereafter named BGP). 
The disadvantage of the single photon mode is the low  
statistics. In the case of the TPOL, a $10\,{\rm W}$ laser with green 
light is used, corresponding to $n_\gamma\approx 0.01$ per bunch. 

The multi-photon mode becomes advantageous when the background is large. 
The LPOL uses a high energy pulsed laser and typically a thousand photons 
per bunch are produced per laser-beam interaction.  
This corresponds to about $10\,{\rm TeV}$ energy measured in the calorimeter. 
The disadvantage is thus related to the energy linearity of  
the calorimeter which is only calibrated at low energy.  
The statistics are also limited by the laser pulse frequency 
($100\,{\rm Hz}$ for the LPOL). 

With respect to these two modes, the determination of 
the longitudinal polarisation in the few photon mode is more involved 
since one has to consider a Poissonian superposition of BCPs resulting in 
a complex energy spectrum (one observes the total energy in the calorimeter 
which can be made of many combinations of few BCPs as well as background 
photons). The BCP energy spectrum is thus given by 
multiple convolutions of the one BCP energy distribution of 
Eq.(\ref{Compton-xsec}) and
therefore becomes a non-linear function of the electron beam polarisation.
However, the advantage of the few photon mode is threefold:  
firstly the statistics are larger, secondly the signal 
over background ratio is higher than that of the single photon mode 
and finally the multiple kinematic edges,
which are independent of the electron beam polarisation,  
can be used to calibrate the calorimeter and determine eventually 
the non-linearity of the energy measurement. 

The main result of this article is the experimental demonstration of 
high precision
measurement of the HERA electron beam polarisation in the few photon mode.

\section{Experimental Setup} 
\label{sec:cavity} 

In order to reach the few photon mode with the given HERA electron beam 
current, one has to use a continuous laser beam of a few kilo Watt power. 
At the time when the experiment was designed (1999-2000), 
the only way\footnote{Indeed, neither the very high power fiber 
lasers~\cite{limpert} nor the locking of mode locked (pulsed) laser beams to 
high finesse Fabry-Perot cavities~\cite{jones}
were available or, at least, reasonably conceivable.} 
to reach such a high power was to use a Fabry-Perot optical 
resonator~\cite{kogelnik,siegman} fed by a continuous laser beam.

Fabry-Perot cavities are a widely used optical device. Since their properties 
are described in many textbooks (e.g.~\cite{siegman}), only their main
characteristics are mentioned here. The cavity of our experimental setup 
consists of two identical spherical mirrors
of very high reflectivities $R>0.999$, where $R$ is the reflection 
coefficient for the beam intensity. Now, considering the simple case of 
an optical plane wave, an optical resonance occurs when the length 
between the two mirrors $L$ is equal to an integer number, $q$, 
times the laser beam wavelength $\lambda$.
Denoting the laser beam frequency by $\nu_L=c/\lambda$,
this condition reads $\nu_L=q\times {\rm FSR}$ where ${\rm FSR}=c/(2L)$ is 
the cavity Free Spectral Range.
Small corrections to this resonance condition appear~\cite{kogelnik} 
when one considers the eigenmodes of the resonator. 
But in any case, when this condition is fulfilled, 
the incident laser beam power $P_{\rm laser}$ is enhanced inside the cavity 
by a factor $F/\pi$, where $F=\pi \sqrt{R}/(1-R)$ is the cavity finesse. 
However since the resonance Full Width at Half Maximum (${\rm FWHM}$) is 
defined by ${\rm FWHM}={\rm FSR}/F$, the highest $F$ (and hence
the highest enhancement factor) leads thus to the smallest ${\rm FWHM}$. 
This means that in order to keep the system at resonnance, one must control 
the value of $\nu_L$ with a relative precision better than 
${\rm FWHM}/(q\times {\rm FSR})$. For the $2\,{\rm m}$ long cavity and 
Nd:YAG laser ($\lambda=1\,064\,{\rm nm}$) used in the cavity polarimeter and
described in the next section,
one gets ${\rm FWHM}/\nu_L\approx10^{-11}$.
This is the precision needed for matching the cavity length and 
the laser beam wavelength to maintain the cavity in resonance.
This is done with a feedback system acting on the laser 
beam frequency, and by using a monolithic cavity.

In practice, the cavity enhancement factor can be reduced by three main 
sources: an error in the mode matching of the transverse mode of the laser to
the fundamental Gaussian mode of the cavity~\cite{siegman} 
(which can be induced, for example, by misalignments~\cite{anderson}); 
an error in the frequency feedback (which can happen, for example,
because of a wrong estimation of the feedback bandwidth with respect to 
the laser or cavity frequency noise density spectra); some differences 
between the optical properties of the two cavity mirrors. 
The measured coupling between the laser beam and the cavity was about $70\%$ 
during the operation in the HERA tunnel.

\subsection{Design and Realisation of the Cavity for HERA}\label{sec:design}

The experimental setup of the cavity polarimeter consists of a laser 
and other optical components inside or close to a Fabry-Perot cavity, 
built around the HERA electron beam pipe.
This provides circularly polarised photons which interact with 
electrons from the main beam of the HERA collider, about 100 meters away 
from the HERMES experiment. This setup is complemented by a calorimeter,  
about $60$ meters downstream of the cavity, for detecting and measuring
the BCPs.

\begin{figure}[htbp]
\begin{center}
\includegraphics[width=\textwidth]{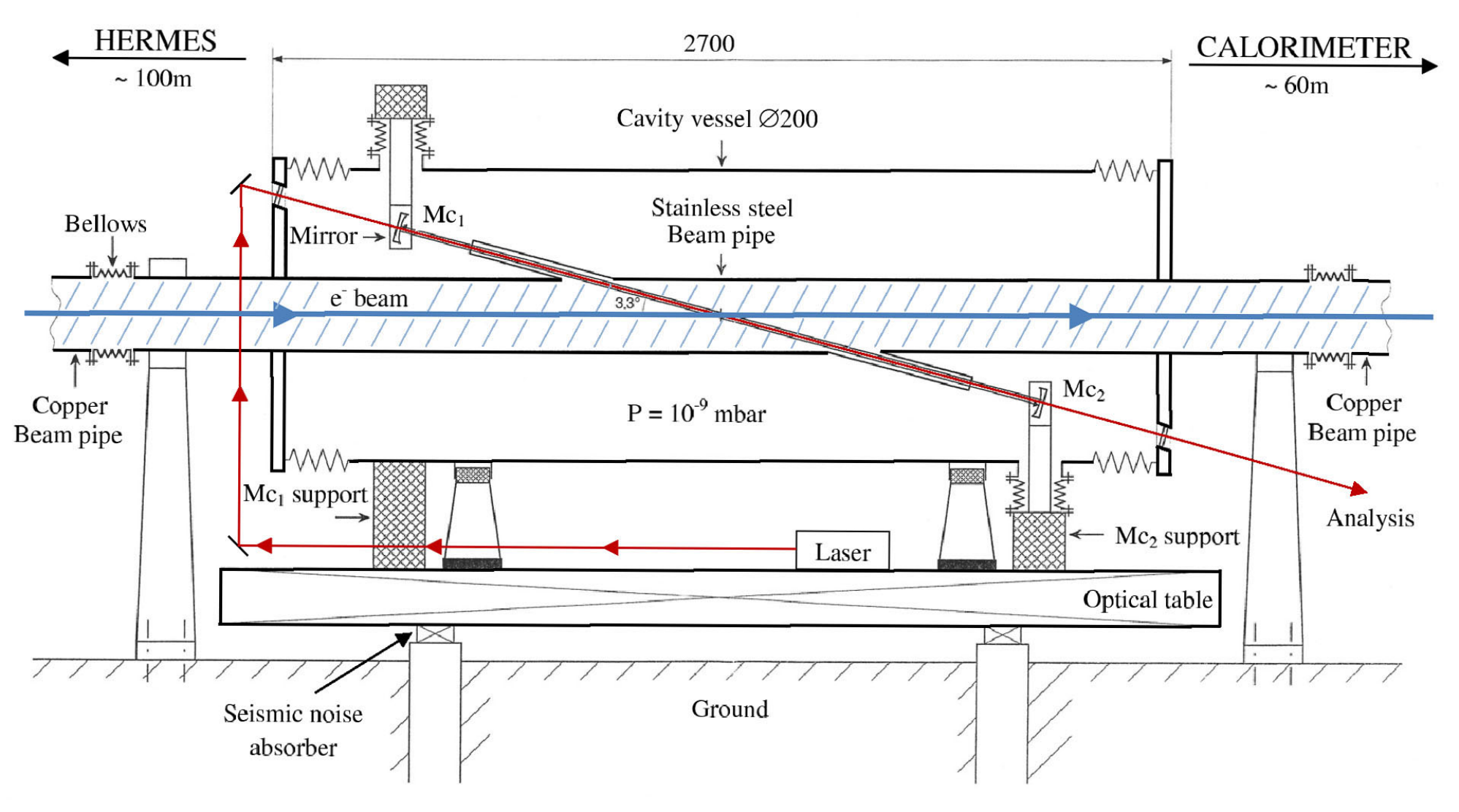} 
\end{center}
\vspace{-3mm}
\caption{Scheme of the cavity surrounding the electron beam pipe with the
laser and main mirrors.} 
\label{fig-drawing-cavity} 
\end{figure}

A schematic view of the monolithic cylindrical cavity vessel, 
with its two mirrors around the electron beam pipe separated by 
a distance of about $2\,{\rm m}$ from each other,
is illustrated in Fig.~\ref{fig-drawing-cavity}.  
The cavity mirrors, with a $2\,{\rm m}$ curvature radius
(confocal cavity~\cite{siegman}, the most stable from the mechanical point of 
view), were installed as close as possible to the beam pipe, 
to minimize the electron-laser beam crossing angle ($3.3$\,degrees).
The laser and all other optical components
were located on an optical table outside of the cavity. 

For installing such a cavity around a ring accelerator beam pipe, 
two main difficulties arise: the presence of the
wake field from the electron beam 
and the presence of vibrations in the whole tunnel environment.
The wake field from the circulating electron beam should not  
disturb the cavity operation, and the  
cavity should not affect the electron beam.  
The propagation into the cavity of high frequency modes  
from the passing beam is suppressed by introducing 
two $15\,{\rm mm}$ diameter metallic tubes around the laser beam extending 
to a length of $\pm 80\,{\rm cm}$ from either side of the holes in the beam 
pipe.
A simulation has shown that this reduces the electron beam
power loss through electromagnetic heating to  
an almost negligible level ($18\,{\rm W}$ during injection,  
less than $0.1\,{\rm W}$ during normal beam operation~\footnote{Estimated by 
S.~Wipf from the DESY accelerator group.}).
Optical calculations, further confirmed experimentally,
have also shown that the tubes do not perturbate the cavity resonance conditions.

\begin{figure}[htbp]
\begin{center}
\includegraphics[width=\textwidth]{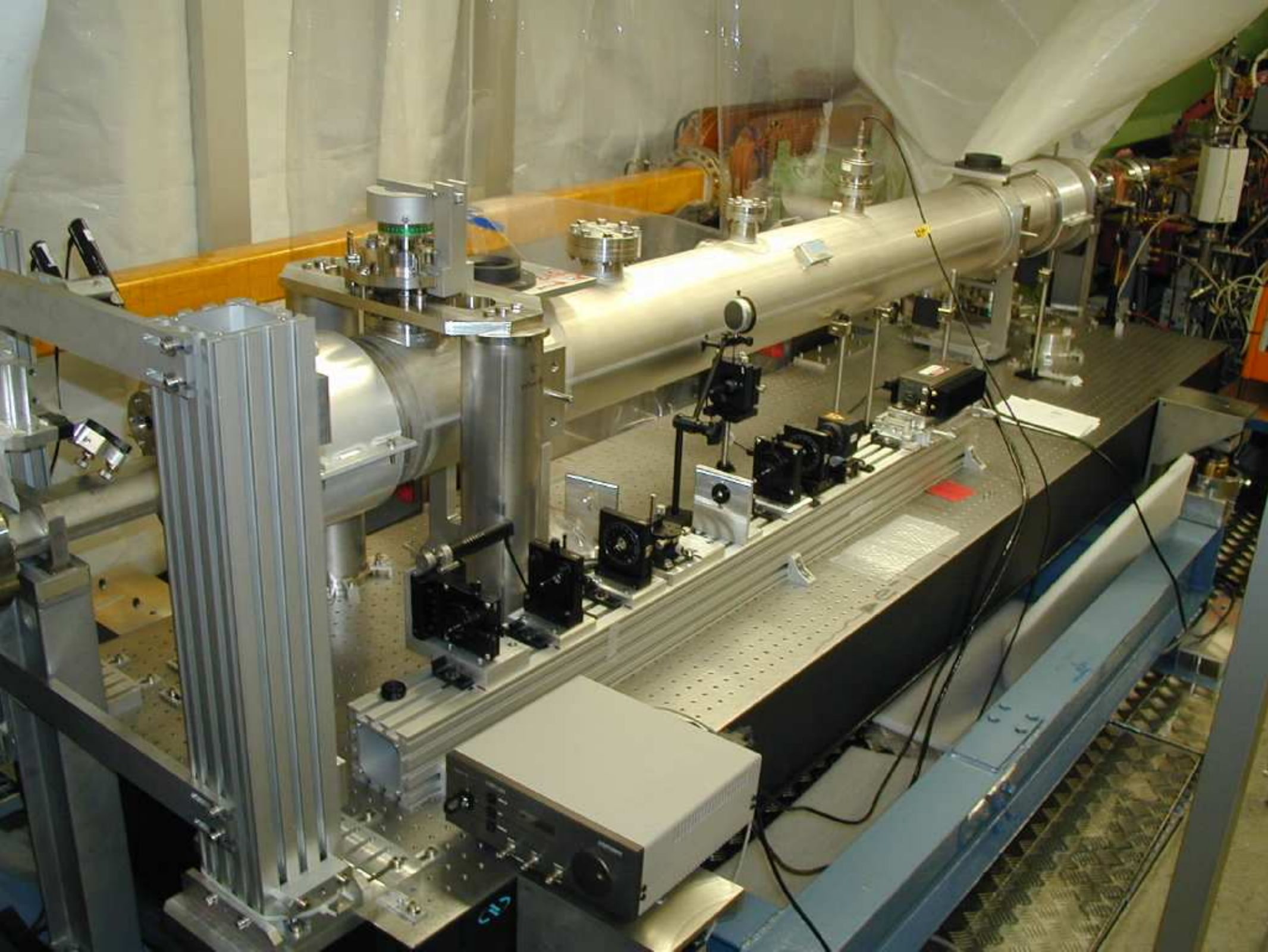}
\end{center}
\caption{Picture of the cavity taken during the installation.  
The laser and the optical elements before the cavity entrance are located  
on the rail parallel to the cavity vessel.}
\label{fig-cavity-photo} 
\end{figure}

Another important requirement of the whole device is a very good 
mechanical stability with time.
This means essentially no vibrations and a small temperature variation. 
The overall realisation is shown in Fig.~\ref{fig-cavity-photo}. 
The whole optical setup, including the cavity mirrors, is vibration isolated: 
the beam pipe inside the cavity vessel, attached to the cavity end-flanges,  
is isolated from the rest of the beam pipe by two standard HERA bellows  
sitting outside the cavity,  
and from the cavity vessel by two other big bellows;
% part of the cavity vessel; 
the optical table feet are equipped with elastomer isolators  
to cut vibrations from the tunnel ground,  
and passive elastomer absorbers to isolate the table from all items  
in contact (cavity vessel, vacuum pumps etc);  the mirror mounts are rigidly
clipped on the optical table and
are linked to the vessel through metal bellows  
thus filtering the remaining vibrations.  
In this way, the cavity mirror holders are completely part of the optical table  
which supports all the optics. The mirror mounts  are able to be  
rotated  manually in all directions while keeping the mirror centre position fixed.

To control the thermal expansions of the cavity and of the optical table,  
the whole system is surrounded by an isotherm house shielded by a
$3\,{\rm mm}$ lead sheet to protect the system against synchrotron radiation.
In addition, a thick lead protection has been installed around the beam pipe  
outside the cavity, and in front of the cavity.  
The laser has been surrounded by a big mu-metal, steel and lead bunker. 
Inside this house, the temperature is controlled and kept constant to well 
within $\pm1^\circ\,{\rm C}$ via heating ribbons and feedback sensors.

\begin{figure}[htbp] 
\includegraphics[width=\textwidth]{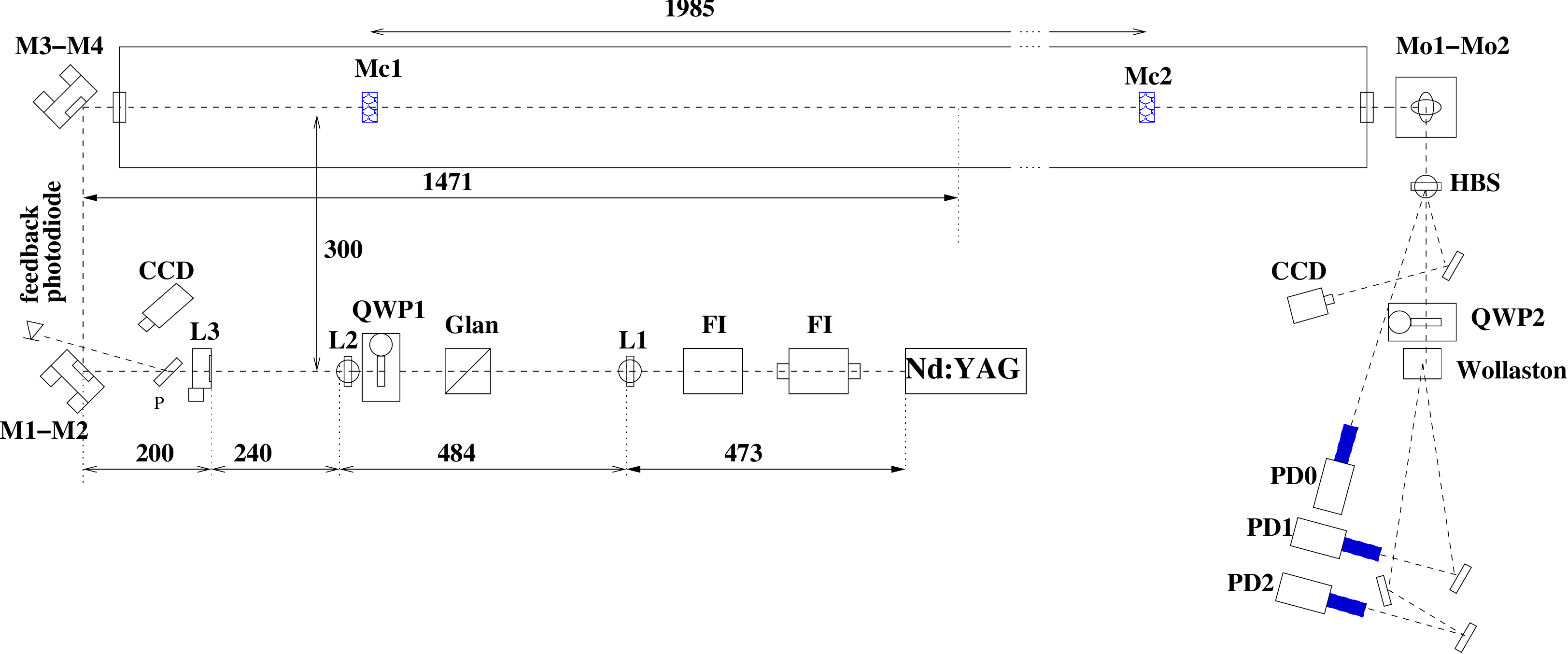}
\caption{Schematic view of the optical scheme. The scales are shown in mm.}
\label{fig-optical-setup} 
\end{figure}

A schematic view of the optical scheme is shown in  
Fig.~\ref{fig-optical-setup}. 
Apart from the cavity, there are two main parts: the entrance for 
providing the laser beam in the cavity and the output for 
measuring the polarisation of the laser beam. 

The laser source is a commercial non-planar ring Nd:YAG oscillator~\cite{byer}
from the Lightwave company. The laser beam frequency can be modified in 
two ways:
\begin{enumerate}
\item 
A piezo-electric transducer is located on the laser  
rod, thereby modifying the rod geometry and therefore the laser beam  
frequency. This is a fast and fine tuning: the laser beam frequency changes 
by $3.4\,{\rm MHz}$ per Volt applied on the actuator within a bandwidth of 
$\approx 30\,{\rm kHz}$. 
\item 
The rod temperature can be varied thanks to a  
Peletier module (controlled by a DC voltage). 
This temperature variation induces a change of the laser beam frequency of  
$5\,{\rm GHz}$ per Volt applied on the Peletier module. This is a  
slow frequency variation with a bandwidth of $\sim 1\,{\rm Hz}$.  
\end{enumerate}

The optical scheme at the entrance of the cavity is designed to 
fulfill the five different functions: 
(a) isolation of the laser oscillator, with a double stage Faraday Isolator 
(FI), a Glan-Thomson prism is used to get a purely linearly polarised state;
(b) obtaining a circular polarisation of the laser beam and switching between
   left and right circular polarisation by using a fast and precise remotely
   controlled rotating quarter wave plate (QWP1 or MOCO);
(c) laser/cavity mode matching, using three lenses (L1, L2, L3);
(d) extraction of the cavity-reflected photon beam to be used for the 
    feedback system;
(e) laser/cavity geometrical alignment using four flat $45^\circ$ dielectric
    mirrors M1, M2, M3 and M4.

  The laser beam enters and exits the cavity vessel through vacuum windows.  
The two cavity spherical mirrors (Mc1 and Mc2), with one inch diameter and 
$2\,{\rm m}$ curvature radius, are coated for $1\,064\,{\rm nm}$ wavelength
with several Ta$_2$0$_5/$SiO$_2$ quaterwave stacks.
The coating losses (due to diffusion and absorption) are very small
($\approx 1.5\times 10^{-6}$), and the transmission coefficient amounts to 
$T=1-R\approx 10^{-4}$.
The finesse of the cavity, calculated and measured as $30\,000$, remained   
constant until the end of the data taking period.

An ellipsometer located at the output of the cavity is used to measure 
the laser beam polarisation. It is described in details in 
a companion article~\cite{mj} to the present one.
The principle of the measurement is to send a light beam of any unknown 
polarisation through first the QWP1 and then the cavity, and from there it is
guided with two mirrors (Mo1 and Mo2) to go through a holographic beam sampler 
(HBS) and another quarter wave plate (QWP2). 
By rotating this latter plate, the polarisation state of the light is modified 
and the state at the exit of the plate depends on the state at the entrance. 
A polariser (Wollaston prism) placed behind the plate spatially separates
the beam into two orthogonal linearly polarised states.
The analysis of the intensities of these two beams in photo-detectors (PD1
and PD2 with PD0 being used as a reference), for various azimuthal angles 
of the QWP, allows the deduction of the polarisation of the incident beam.

\subsection{The Calorimeter}\label{calo}  

A sampling calorimeter~\cite{lorenzon00}   
was used to measure the energy spectrum of the back-scattered Compton 
photons as well as photons from background processes. 
It was located on a movable table which can be remotely moved
away from the photon beam line during the beam injection and may be used
to optimise the rate of measured BCPs during the data taking. 
The calorimeter has a total of $24$ layers where each layer consists of  
a $3.0\,{\rm mm}$ tungsten (W) absorber plate  
and a $2.63\,{\rm mm}$ scintillator plate. The plates are 
$40\times 40\,{\rm mm}^2$ and are optically coupled on all four sides to  
wavelength shifter (WLS) 
plates that bring light to one Photo Multiplier Tube (PMT) placed  
at the back of the calorimeter behind a $27\,{\rm mm}$ W shielding plate.  
The PMT transmutes the light from the WLSs into  
an electrical signal.  
The calorimeter response has been 
simulated~\cite{lorenzon00} using the Monte Carlo program 
{\sc Geant}3~\cite{geant321}. 

The energy resolution and uniformity have been measured in test beams 
at DESY.  
The measured energy resolution $\sigma/E=a/\sqrt{E(\rm GeV)}\oplus b$,
with $a=16.0\%$ and $b=0.3\%$, is in agreement with 
the Monte Carlo simulations. 
The uniformity scan showed less than 5\% deviation from linearity for 
up to $\pm 10\,{\rm mm}$ from the calorimeter centre.  
The linearity is measured to about 0.2\% over the full energy range 
of 1 to $6\,{\rm GeV}$ of the test beams.

\subsection{Electronics and DAQ}\label{daq} 

For the polarisation measurement two electronic systems are designed.
The first one is the feedback system which locks the laser frequency on 
one of the Fabry-Perot cavity harmonics, and the second one is the scattered 
photon readout system.
We shall first describe briefly the cavity feedback system and then
concentrate on the fast calorimeter DAQ, the main novel aspect of 
our electronic readout system.

\subsubsection{Feedback System}\label{section-feedback} 
\begin{figure}[htbp]
\begin{center}
\includegraphics[width=0.775\textwidth]{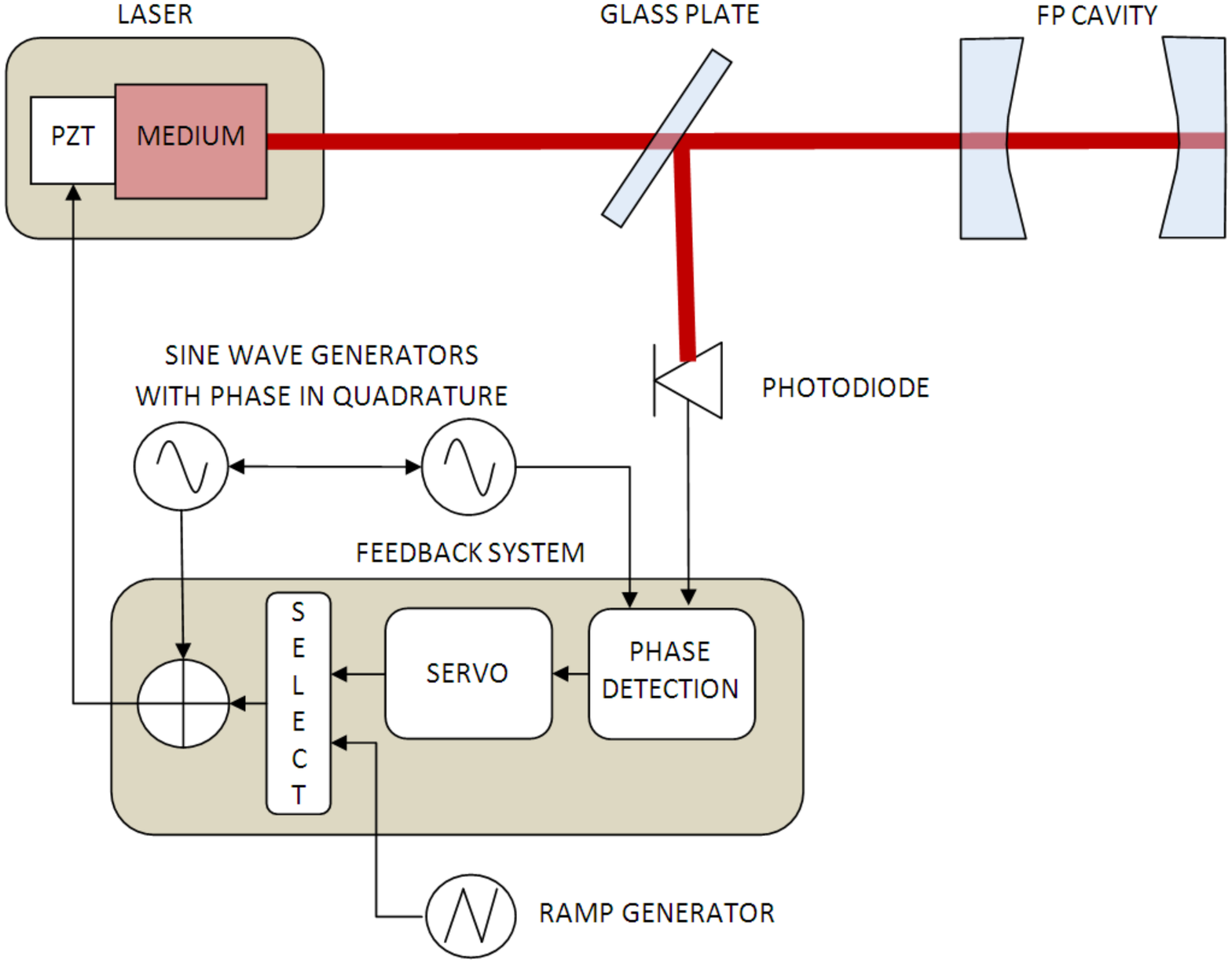}
\end{center}
\vspace{-5mm}
\caption{Simplified view of the feedback system (see text).} 
\label{fig-pound} 
\end{figure}

The feedback system (Fig.~\ref{fig-pound}) is based on the Pound-Drever-Hall 
(PDH) method.
We modulate the incident photon beam at $930\,{\rm kHz}$ in order to 
create two sidebands. The modulation frequency is much higher than
the bandwith of the cavity. When the main laser frequency is near 
a Fabry-Perot harmonic, the main part of the photon beam is absorbed but
the sidebands are reflected. This reflected beam is processed to obtain
the error signal for the feedback.

The feedback system works in 2 steps:
\begin{itemize}
\item
The first step is to find a resonance frequency. The system selects the
ramp generator which is applied to the laser PZT. The laser frequency
shifts slowly and can cross one of the cavity resonance frequencies.
\item
Once a resonance frequency is reached, the feedback system switches
and closes the feedback loop. An analog filter of many orders allows to
maintain the cavity resonance.
\end{itemize}

\subsubsection{Calorimeter Fast DAQ System}\label{seccalo}
\begin{figure}[htbp] 
\begin{center}
\includegraphics[width=0.825\textwidth]{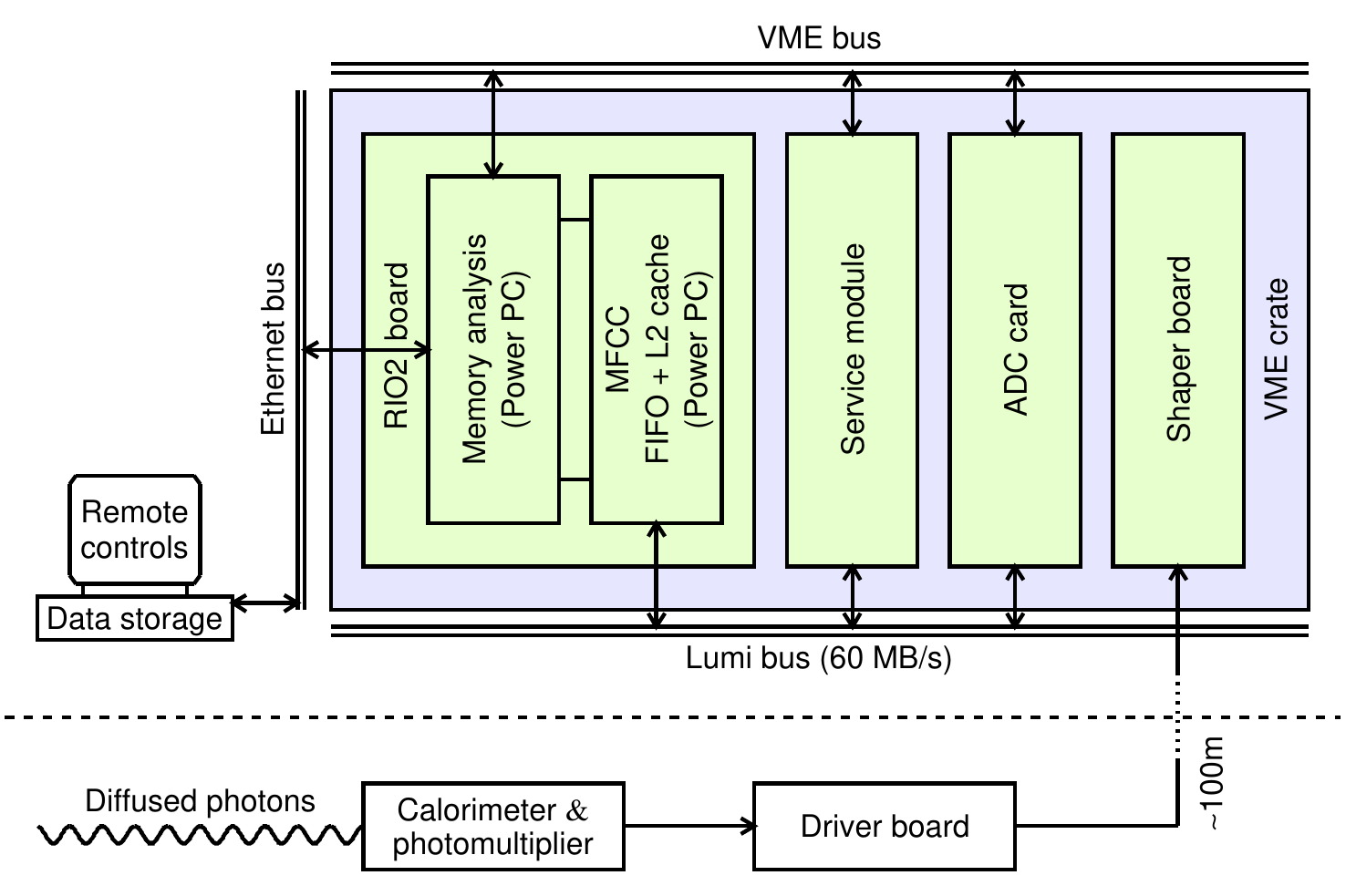}
\end{center}
\vspace{-3mm}
\caption{Global architecture of the DAQ system.} 
\label{fig-rio2} 
\end{figure}

The DAQ hardware components (Fig.~\ref{fig-rio2}) 
are similar to those used for the electronic 
upgrade of the HERA transverse polarimeter and for the new H1 
luminometer~\cite{lumi}.
The main difference lies in the fact that the cavity polarimeter does not use  
any trigger to reduce the event rate. Therefore the DAQ system has to be fast  
enough to cope with the HERA bunch frequency of $10.4\,{\rm MHz}$. 

The energy spectrum of the emitted photons is measured in terms of charges 
collected in the tungsten-scintillator sampling calorimeter and read out 
by a photomultiplier. A driver board is used to amplify the signal before 
being sent along $\sim 100\,{\rm m}$ of cables to a shaper board. 
A dummy channel is also sent along the cables for the subtraction of 
the common noise by the shaper board. Due to the long cable length, 
the signal has very long tails due to the skin effect, thus the signal is 
optimized to have the much longer signals in $96\,{\rm ns}$ 
($10.4\,{\rm MHz}$) to reduce sensitivity to jitter by overdriving some high 
frequencies.

The shaped output is fed to an ADC card in the form of a differential signal 
up to $2\,{\rm V}$, with a return to the baseline below $1\%$ in 
$96\,{\rm ns}$. The fast shaping is essential to avoid electronic pileup 
from one bunch to the next.

The digital part of the acquisition system fits in a standard $22$-slot 
VME crate. It includes the ADC card, a service module which drives all 
clocks in the system and a commercial Power-PC processor board composed 
of a VME mother board RIO$2$ $8062$~\footnote{RIO stands for Reduced 
instruction set computer I/O, RIO2 8062 board has a 603r Power-PC of 
$300\,{\rm MHz}$.} and a PCI (Peripheral Component
Interconnect) connected daughter board MFCC $8442$~\footnote{Multi-Function 
Computing Core card 8442 contains a 7400 
Power-PC of $466\,{\rm MHz}$, a $128\,{\rm Mbyte}$ SDRAM (Synchronous Dynamic 
Random Access Memory), a 64-bis Power-PC bus of $66\,{\rm MHz}$, 
a $1\,{\rm Mbyte}$ level 2 cache and two fast FPGA (Field-Programmable Gate 
Array) of Altera 10K50.}, both from CES~\cite{ces}. 
A dedicated $14$-slot wide fast readout bus backplane is used to transfer 
signals between these boards.

The ADC card digitizes the analog signal from the shaper board in $12$ bits 
at $41.6\,{\rm MHz}$ and stores $4$ samples per bunch crossing 
in $2$ independent pipelines of a depth of $512$ samples each. 
One is readable via the VME bus and the other via the fast readout bus.

The service module receives the HERA clock ($96\,{\rm ns}$) and the machine 
cycle (first bunch) signal. It provides two internal clocks, 
which are $2$ and $4$ times faster than the HERA clock, 
to drive the ADC sampling clocks and 
does the phase adjustment between the ADC samples and HERA clock. 
The phase is adjusted such that the maximum signal corresponds to 
the second ADC sample and the baseline to the fourth sample. 
The difference of the two thus gives the signal amplitude.
Note that, since the 
fast readout bus transfers two ADC samples together in $24$ bits, there is an 
ambiguity in the second and fourth ADC samples. Due to this, the maximum can 
be either in the second sample and the baseline in the fourth one or vice 
versa, depending on the exact moment when the DAQ program is started.

The readout sequence and bus protocol are controlled by a FPGA  
(Field-Programmable Gate Array) located on the MFCC board.
The core of the fast acquisition program runs into the MFCC Power-PC 
attached to  
the RIO$2$ Power-PC and consists of two nested loops. The outer loop runs on  
a requested number of HERA turns (default is $400\,000$), the inner loop runs  
on the $220$ bunches with $4$ times $12$-bit samples per bunch for one HERA  
turn. 

The data from the ADC pipeline is added to the FIFO (First In First Out) of  
the front-end FPGA of the MFCC board through the fast readout bus in $24$ bits  
every $20.8\,{\rm MHz}$. The data in the FIFO are packed in a $64$-bit format  
($4$ times $12$ bits plus status bits) which corresponds to $4$ ADC samples  
($1$ bunch).  A block of $4\times 64$-bit FIFO data is then transferred  
through the $64$-bit Power-PC bus at $66\,{\rm MHz}$ to a $1$-Mbytes level $2$ 
(L$2$) cache of the MFCC. 
A whole HERA beam turn of $220$ bunches needs thus $55$ of such  
transfers. In the cache, for every one of $220$ bunches, a histogram is  
created and filled using by default the difference of the second and  
fourth ADC samples. 

The L$2$ cache may accommodate $220$ histograms each containing $1024$ bins 
with a maximum bin content of $65535$ ($16$ bits). 
The gain of the analog chain is chosen in order to get a full dynamic range 
of about $90\,{\rm GeV}$. 
The histogram bins are then scaled down to $512$ bins whereas the bin content  
is expanded to $32$ bits (maximum content is increased to $4\times 10^9$)  
to avoid potential overflow in the bin content. In an improved version of  
the histogramming code, three variable bin sizes are further used to store  
in an optimal way the energy distribution for the given $512$ bins:  
finer bin size at low energies which contain both the high statistics and  
the BCP signals and two coarser bin sizes at medium and high energies. 

Since there is no hardware synchronisation between the HERA clock and  
the histogram filling, a system of waiting loops is implemented in the  
acquisition code to perform the matching and to avoid full or empty FIFO  
issues. Once the requested amount of data ($400\,000$ HERA turns) is reached,  
the histograms are transferred to the Power-PC of the RIO  
board via a PCI bus and the memory is refreshed. 
The transfer takes less than $100\,{\rm ms}$ before  
another acquisition cycle can start. Two successive cycles correspond to  
two different circular laser polarisations; one left handed and the other  
right handed. The matching between a given laser polarisation state and  
the corresponding histograms is ensured in the acquisition code. 

The histogram data are sent through the local network into a dedicated PC 
for online processing and publishing of results on the DESY network, 
as well as for storage for subsequent offline analysis.
On the same PC, the DAQ program can be started and stopped remotely from any  
other PC connected to it.

%\newpage
\section{Determination of the Electron Beam Polarisation}\label{sec:method}

\subsection{Principle of the Analysis Method}

The available data for the polarisation measurement of an individual 
bunch consists of a pair of photon energy histograms each with $512$ bins.
These are successively recorded by the calorimeter DAQ
for each bunch during one DAQ period ($\approx 10\,{\rm s}$)
for the two polarisation states
of the laser beam $S_3=+1$ and $S_3=-1$. These spectra come from
a sum of genuine BCPs from electron-laser
interactions, and three main backgrounds: the already mentioned BGP, 
the electron beam scattering off
Black Body Photons (BBP) emitted by the hot beam pipe and
the Synchrotron Radiation Photons (SRP).

Each histogram provided by the calorimeter DAQ is thus a set of
integer numbers $\{h_i\}_{i=1,...,512}$ in a measured energy
interval $[E_{i-1},E_i]$ for a given bunch.
Since the DAQ operates without trigger, the total number of 
entries of the histograms are fixed to the
number of HERA turns $N_{\rm loops}$ accumulated during one
DAQ period $\sum_i h_i=N_{\rm loops}=400\,000$.
This constraint is necessary if one wants to extract precisely
the electron beam polarisation for all bunches
from a fit to a pair of two histograms corresponding to 
$S_3=+1$ and $S_3=-1$ laser polarisation states in the few photon mode.

The numerical
procedure that has been set up to extract the polarisation from the 
bunch energy histograms is the following:
\begin{itemize}
\item
The theoretical energy spectra of one BCP (see Eq.(\ref{eq:bcp})), 
BGP and BBP~\cite{zomer} are computed numerically (the SRP background
is considered as a pedestal and it is treated separately as described below).
\item
All these spectra are mixed to provide an energy histogram
according to the method described below.
\item
The detector effects (calorimeter, ADC/energy conversion and electronic noise)
are applied to the energy histograms.
\item
A comparison between the experimental and calculated energy histograms
is performed using a likelihood fit. The unknown parameters of this fit are 
the luminosities of the BCP, BGP and BBP processes integrated
over one DAQ period, the electron beam 
polarisation and  the parameters describing the detector effects.
\end{itemize}
It was verified numerically that, as mentioned in Sect.\ref{sec:meas_mode},
the simultaneous determination of the electron
beam polarisation and the other parameters can be obtained 
since the energy histograms possess kinematic edges independent of 
the electron and laser beam polarisations. 

A detailed derivation of the formula used in our fits can be found
in \cite{zomer}. Here we just indicate the main ingredients of the
method. This analysis method uses probabilities
which are discretised.
The probability $p_i$ for a photon to be in the energy bin $i$
is computed from the theoretical differential cross sections
and includes the following independent contributions:
\begin{equation}\label{fragmentation}
p_i=a_{{\rm BCP}0}\times p_{i,{\rm BCP}0}+ a_{\rm BCP1}  \times p_{i,{\rm BCP}1}+a_{\rm BGP}
\times p_{i,{\rm BGP}}
+a_{\rm BBP} \times p_{i,{\rm BBP}}
\end{equation}
where $\sum_i p_{i,x}=1$ with $x$ representing the different processes, 
$a_x$ depend on the luminosities of the corresponding 
processes and BCP$0$ and BCP$1$
are related to the BCP differential cross sections, independent of and
linearly dependent on the product $S_3P_z$ respectively
\begin{equation}
\frac{d\sigma_{\rm BCP}}{dE_{\rm BCP}}=
\frac{d\sigma_{{\rm BCP}0}}{dE_{\rm BCP}}+S_3P_z\frac{d\sigma_{{\rm BCP}1}}{dE_{\rm BCP}} \label{eq:bcp}
\end{equation} 
where the energy spectra $d\sigma_{{\rm BCP}0}/dE_{\rm BCP}$ and 
$d\sigma_{{\rm BCP}1}/dE_{\rm BCP}$
are derived from  Eq.(\ref{Compton-xsec}).

The expression for the
total (theoretical) energy follows the Poissonian law and
can be written as:
\begin{equation}\label{energy-law}
P_i = \sum_{N=0}^\infty e^{-M}\frac{M^N}{N !}
\delta\!\left(E_i-\sum_{k=0}^N E_{i_k}\right)
\sum_{i_1}p_{i_1}...\sum_{i_k}p_{i_k}...\sum_{i_N}p_{i_N}  
\end{equation}
where $M=a_{{\rm BCP}0}+ a_{{\rm BCP}1}+a_{\rm BGP}+a_{\rm BBP}$
is the average number of expected photons entering
the calorimeter per bunch crossing (the SRP being treated separately); $N$
is the number of scattered photons (in practice it turns out that 
$N\leq 5$ is enough in our dynamical regime);
indices $i_k$ run over all the bins
of the theoretical histogram. 

Note that Eq.(\ref{energy-law}) is indeed the 
sum of the discrete $N^{\rm th}$ convolutions of the energy histograms so that
one can identify the contribution $N=0$, $N=1$, $N=2$, ... $N=5$ to the 
zero-photon (i.e.\ the SRP peak), one-photon,
two-photon, ..., five-photon energy spectra.  

The probability to get a histogram $\{h_i\}_{i=1,...,512}$ is finally given by:
\begin{equation}\rp =\prod_i (P_i)^{h_i}
\end{equation} 
from which the likelihood estimator may be
extracted: $\rw =-2\ln(\rp)$. 
The likelihood minimization procedure
is standard (see \cite{zomer} for further details).

If the SRP background would have been included in the above formula,
the number
of terms in Eq.(\ref{energy-law}) would have been very large
(from $500$ to $15\,0000$ photons with the critical energy of 
$\approx 40\,{\rm keV}$ and the peak energy
ranging from $20\,{\rm MeV}$ to $6\,{\rm GeV}$).
The adopted solution is to introduce a term in Eq.(\ref{energy-law}) 
that combines (adds) the SRP distributions into a so-called (Gaussian) 
radiation peak, as described in Sect.~\ref{sec:detector-effects}.
The result shows a very narrow distribution in the energy region where
no other photon is present.

\subsection{Detector Effects} \label{sec:detector-effects}
The detector response effects that are taken into account are not explicitly 
shown in the probabilities of Eq.(\ref{energy-law}). Indeed, the calorimeter
has been simulated with {\sc Geant}3~\cite{geant321} in order to derive
a parameterisation describing the energy response. In addition
to the calorimeter response to BCP or BGP photons, a specific study 
has also been devoted to the SRP which corresponds to, as indicated in the
previous section, a large number of low energy photons. 
From the simulation studies, it turns out that the calorimeter response 
function $f_E(E,E_0)$, relating the energy $E$ observed in the calorimeter 
to the `true' incident energy $E_0$,
can be modeled by the following quasi Gaussian distribution  
\begin{equation}\label{wash}
f(E,E_0)dE=\frac{e^{-x}x^{\delta-1}}{\Gamma(\delta)}dx
\end{equation}
where $E= \alpha x^\mu E_0$, $\alpha$ and $\delta$ are given by
$\left<E/E_0\right>=\alpha \delta$ and $\sigma_E^2=\alpha^2 \delta$, and
$\Gamma$ is the special Gamma function.
Note that such a distribution approximates to a Gaussian if
$E_0/\sigma \rightarrow 0$ and vanishes when $E \le 0$, which is not
the case for a true Gaussian.
For $\mu=1$ it also shares with a Gaussian the calorimeter additive property
$\sigma_{E_1}^2+\sigma_{E_2}^2=\sigma_{E_1+E_2}^2 $.
However a more versatile ansatz with $\mu \ne 1$ has been used to better 
approach the simulation especially at low $E_0$.
The functional forms of $\mu(E_0)$ and $\alpha(E_0)$ have been derived from the
simulation studies. They depend on five free detector parameters which are 
determined from the likelihood fit to the data.

Finally, the conversion of the number of ADC counts $N_{\rm ADC}$ to energy is
described by 
\begin{equation}\displaystyle
N^i_{\rm ADC}\propto \frac{E_i/r}{1+\frac{d_{\rm bias}}{E_i}
+s_{\rm bias}E_i} \label{E_ADC}
\end{equation}
where $r$ stands for an energy response factor which may be time dependent
and $E_i$ is the energy in histogram bin $i$,
$d_{\rm bias}$ takes into account the fact that the calorimeter is 
less efficient for energies below $1\,{\rm GeV}$ (the simulation gives 
$d_{bias}=0.035\,{\rm GeV}$), and $s_{bias}$ has been added to take
care of a possible leakage (usually very small) given by a photon beam 
misalignment with respect to the calorimeter center. These three parameters 
are also determined from a fit to the experimental histograms.

Note that other models for the detector response have also been considered 
in order to evaluate the systematic uncertainty related to the choices of
Eqs.(\ref{wash}) and  (\ref{E_ADC}) (see Sect.~\ref{sec:syst}). 

The method developed in the analysis
is to determine the relevant detector parameters 
by minimising the likelihood estimator $\rw$
using the energy spectra selected from some specific 
datasets and bunches
(see Sects.~\ref{sec:condition} and \ref{sec:uc_syst}).  

\begin{figure}[htb]
\begin{center}
\includegraphics[width=0.675\textwidth]{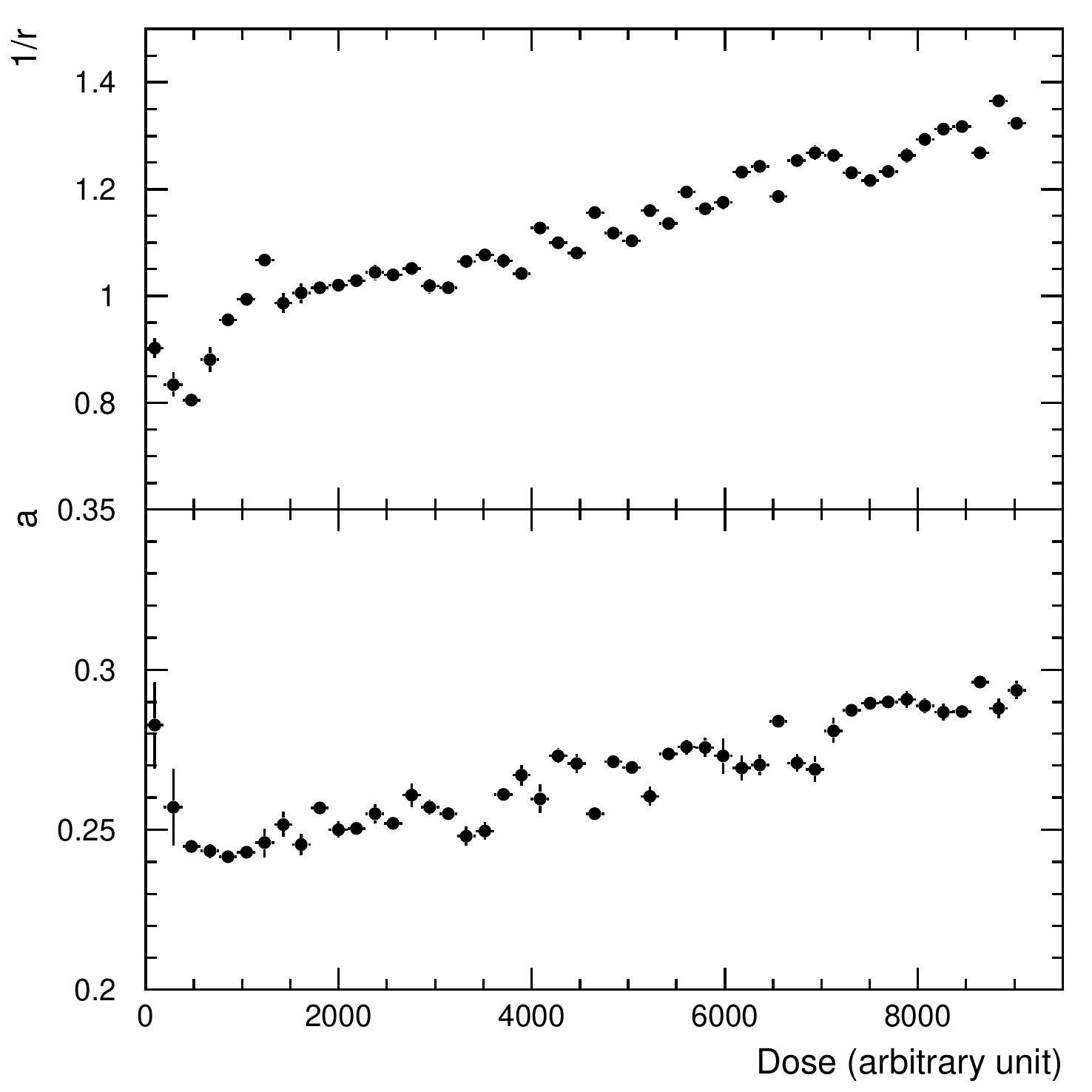}
\end{center}
\vspace{-3mm}
\caption{Observed variation of the calorimeter energy response factor ($r$) 
and resolution (a) as a function of radiation dose in an arbitrary unit.}
\label{fig:gain}
\end{figure}

It was checked numerically that the measured energy spectra allows
an in-situ determination of 
the characteristics of the calorimeter independently of the electron beam
polarisation. This over-constrained feature of the Compton spectra in the 
few photon mode turned out to be crucial for the polarisation measurements,
since the calorimeter could not be calibrated 
with a high energy test beam before its installation and in addition 
it is subject to aging effects. 
Indeed, as it is shown in Fig.~\ref{fig:gain},  
the energy response factor $r$ and resolution $a$ of the calorimeter determined
from the fit to the energy spectra vary
as a function of the radiation dose (which is proportional
to the number of data samples taken by the LPOL cavity). 
The energy resolution $a$ was found to be worse than the test beam result 
(Sect.~\ref{calo}). This is due to the degradation caused by aging effects and 
the addition of a $2X_0$ tungsten plate in front of the calorimeter for
reducing the synchrotron radiation.

%\newpage
\subsection{Comparison of Measured and Theoretical Energy Spectra}
\begin{figure}[htbp]
\begin{center}
\includegraphics[width=0.675\textwidth]{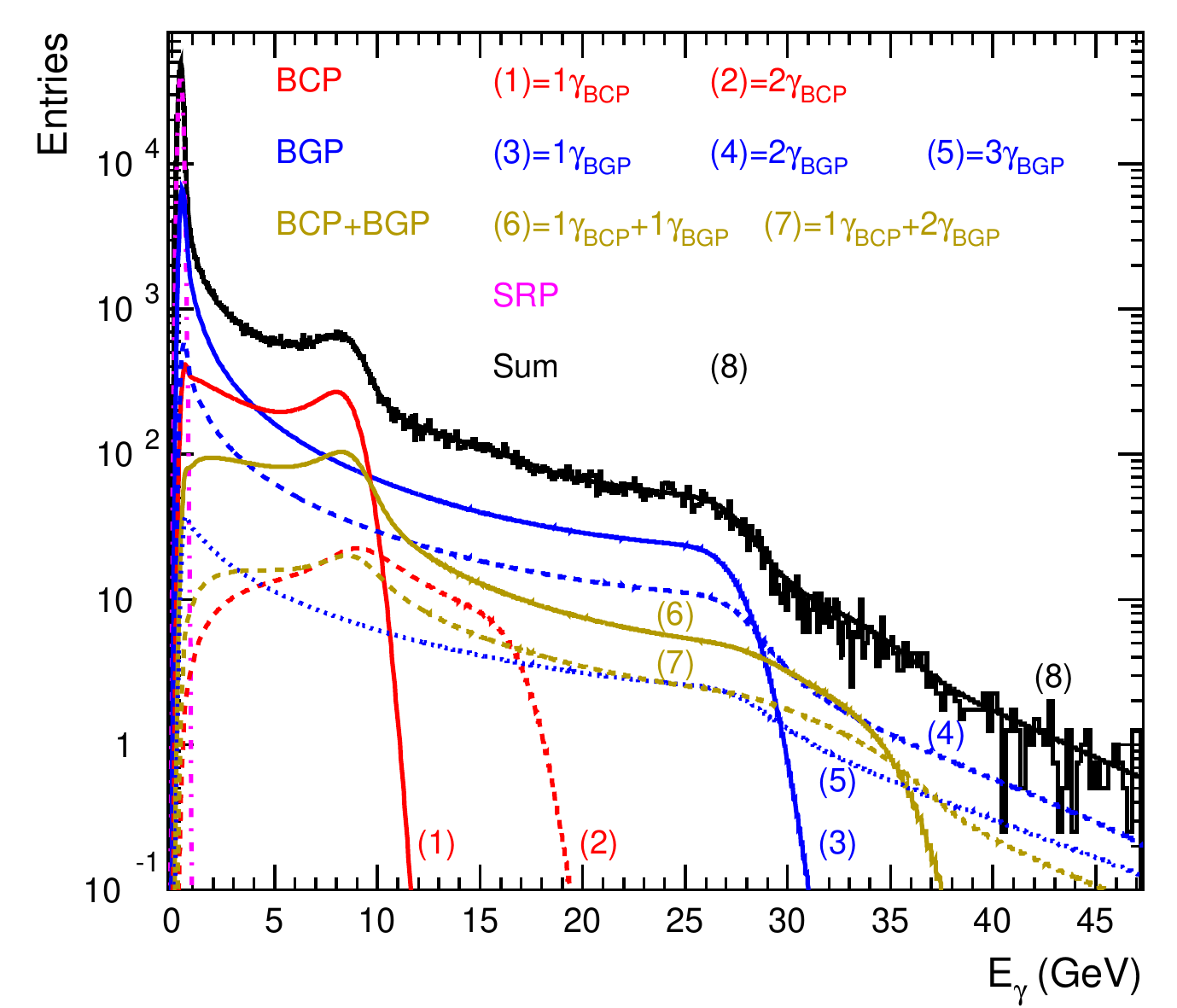}
\end{center}
\vspace{-3mm}
\caption{A measured energy spectrum (histogram) of a given bunch shown
together with different contributions obtained from predictions (curves). 
The data were accumulated during about $10\,{\rm s}$ of acquisition time.}
\label{fig:composant}
\end{figure}

Taking the cross sections for BCP, BGP, and BBP shown in~\cite{zomer},
the whole experimental setup has been modeled. 
The resulting energy spectra 
have been computed for every individual process emitting one or more photons, 
and two processes convoluted together. Finally, the convolution of all 
the processes has been performed. These curves are shown in 
Fig.~\ref{fig:composant} and compared to a measured energy distribution.
An excellent agreement is seen between the data and curve~(8) 
representing the sum of all processes. 
The BBP component is not
shown explicitly but included in the sum in Fig.~\ref{fig:composant}. 
From this typical experimental spectrum, one can clearly see that 
the contribution of three photons is already at the percent level and that 
the two photon contributions reproduce the various kinematic edges which 
appear as small local maxima in the energy spectrum. 
Note that the various contributions are convoluted with the detector response.

\begin{figure}[htbp]
\begin{center}
\includegraphics[width=0.675\textwidth]{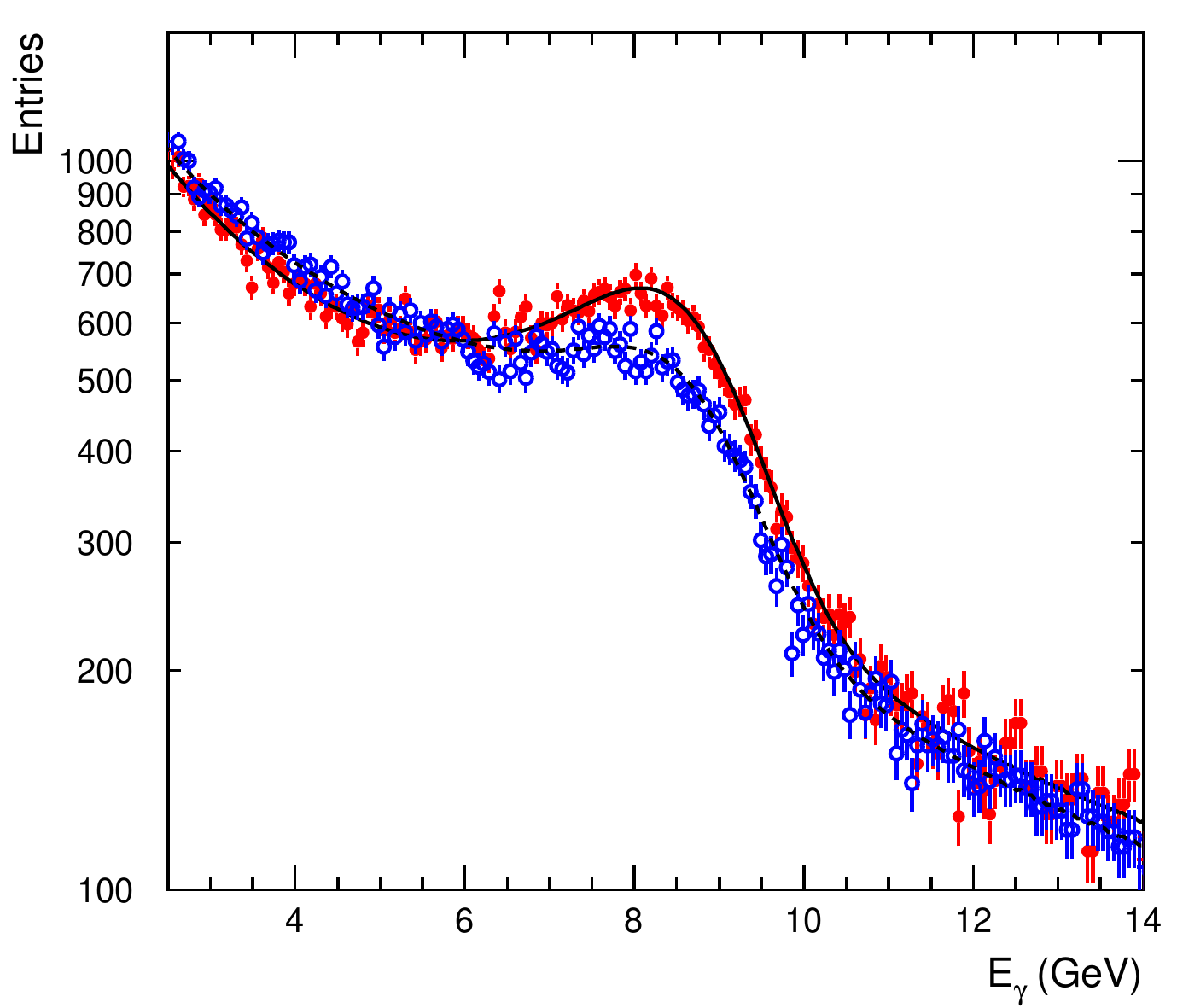}
\end{center}
\vspace{-3mm}
\caption{Experimental energy spectra of a given bunch
corresponding to $S_3=1$ (full dots) and $S_3=-1$ (open dots).
Each spectrum was accumulated during about $10\,{\rm s}$ of acquisition time.
The fit results are also shown (the full and 
dashed curves).}
\label{fig:lr}
\end{figure}

In Fig.~\ref{fig:lr}, the two spectra for laser polarisation 
$S_3=1$ (full dots with the full curve) 
and $S_3=-1$ (open dots with the dashed curve) are shown, 
together with the measured values in the `Compton energy range'. 
The difference allows the electron polarisation measurement, 
as mentioned previously.

\newpage
\section{HERA Running and Cavity Data Taking} \label{sec:condition}
The total LPOL cavity data taking amounts to about $500$\,hours 
(from 6 October 2006 to the end of the HERA running in June 2007).

It is important to consider the operating conditions because they happen
to differ vastly from the foreseen ones. As a consequence the real analysis
method had to be modified or improved in various situations.
\begin{description}
\item{\it HERA beam orbit instabilities:}
Usually the electron beam was very stable, but occasionally it suffered
from rapid changes as can be seen in Fig.~\ref{fig:beam} where the beam 
conditions changed between the $S_3=-1$ and $S_3=+1$ spectra measurements.
Two more parameters were then added to the likelihood estimate to 
account for a different BCP and BGP fluxes in 
the $S_3=-1$ and $S_3=+1$ energy spectra.
In this particular example, the BGP flux has varied by $40\%$ while the BCP
flux remains unchanged. For the polarisation measurement, the variation
of both fluxes is required to be less than $20\%$ which rejects only
a small fraction of spectra (less than $0.4\%$). 
\begin{figure}[htbp]
\begin{center}
\includegraphics[width=.675\textwidth]{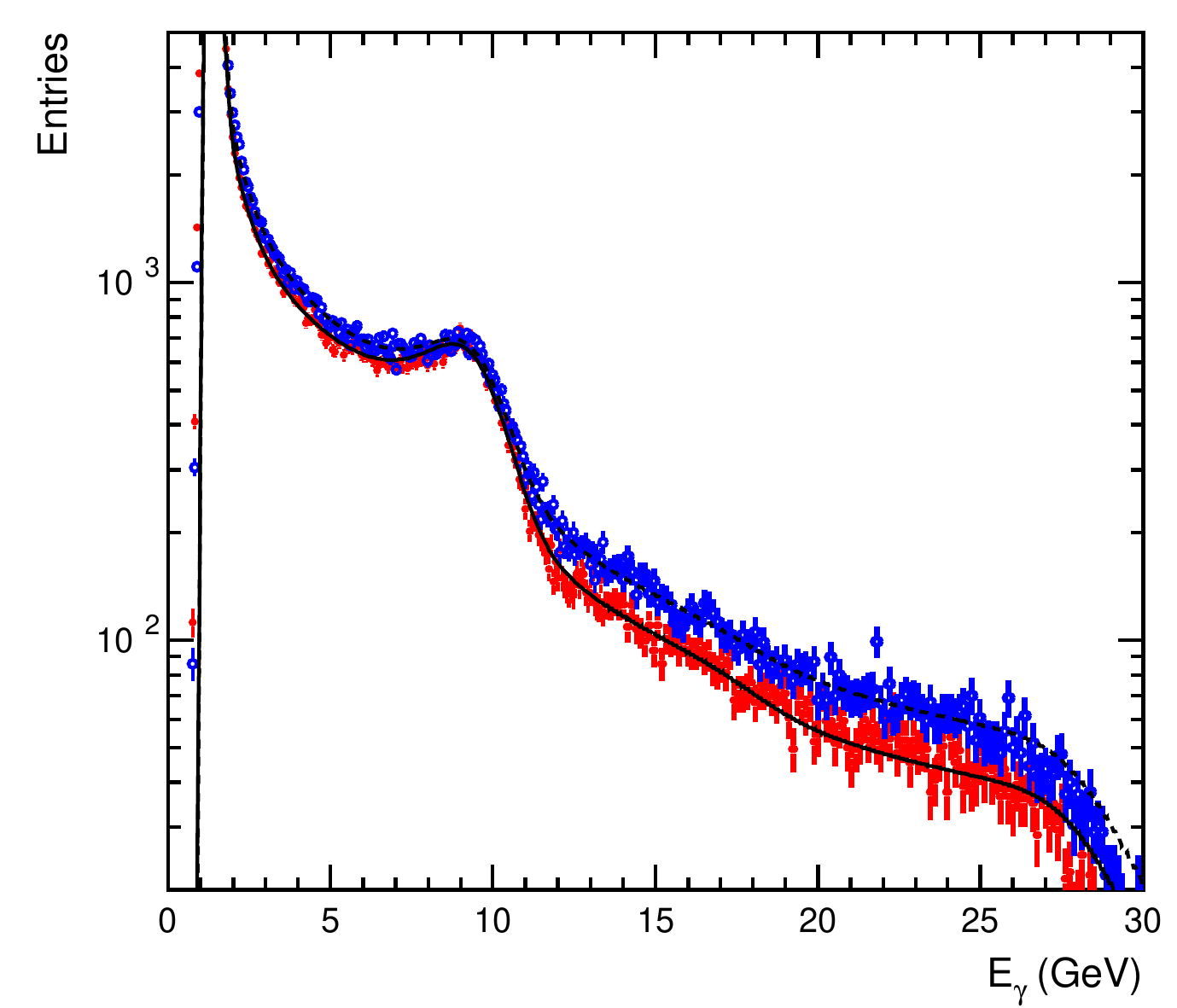}
\end{center}
\vspace{-3mm}
\caption{Energy spectra showing the beam instability between the 
$S_3=-1$ (open dots) and $S_3=+1$ (full dots) spectra measurements.
Each spectrum was accumulated during about $10\,{\rm s}$ of acquisition time.
The curves represent the corresponding fit results.}
\label{fig:beam}
\end{figure}

\item{\it HERA beam orbit slow variations - calorimeter exposure:}
From fill to fill and even during a fill,
the orbit may change and hence so does the impact on the calorimeter
of the BCPs (and BGPs). As the calorimeter is very narrow and 
leakage may happen, this
may induce changes in the detector parameters governing energy scale and
resolution. This is why detector parameters are reestimated on the histograms
themselves by a {\sc Minuit} optimisation procedure every 3 minutes 
($20$ data acquisition samples).
\item{\it Radiative peak position:}
The synchrotron radiation flux, being generated by the magnetic fields 
on the electron trajectory, is sensitive to orbit changes. 
Therefore, one more parameter
is needed to describe its average energy $E_{\rm SRP}$. This parameter
defines the histogram position on the energy scale and is determined 
by a parabolic fit to the maximum of every energy histogram. 
One expects a correlation between the electron beam current and
$E_{\rm SRP}$. The larger $E_{\rm SRP}$
the higher the level of synchrotron radiation. 
This is indeed what we have observed 
(Fig.~\ref{fig:rads}). It should be pointed out that during
the commissioning phase the synchrotron radiation level was found to be 
much higher than was foreseen; this was finally traced back to 
the transverse magnetic field applied to the gaseous fixed target 
in the HERMES experiment from 2001 till the end of 2005.
\begin{figure}[htbp]
\begin{center}
\includegraphics[width=.6\textwidth]{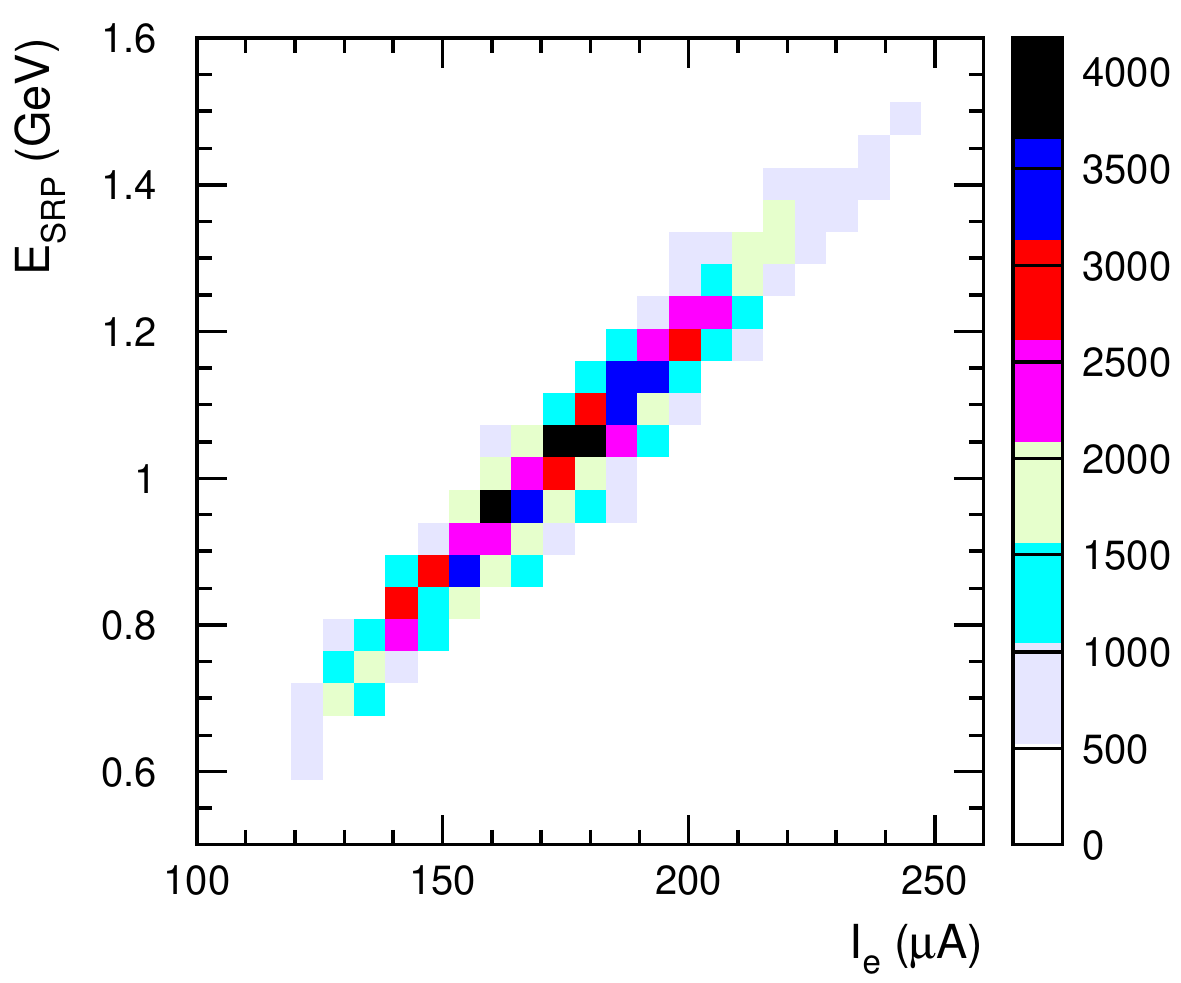}
\end{center}
\vspace{-3mm}
\caption{Observed correlation between the bunch-dependent electron beam 
current and the synchrotron radiation level ($E_{\rm SRP}$).} 
\label{fig:rads}
\end{figure} 

\begin{figure}[htbp]
\begin{center}
\includegraphics[width=0.6\textwidth]{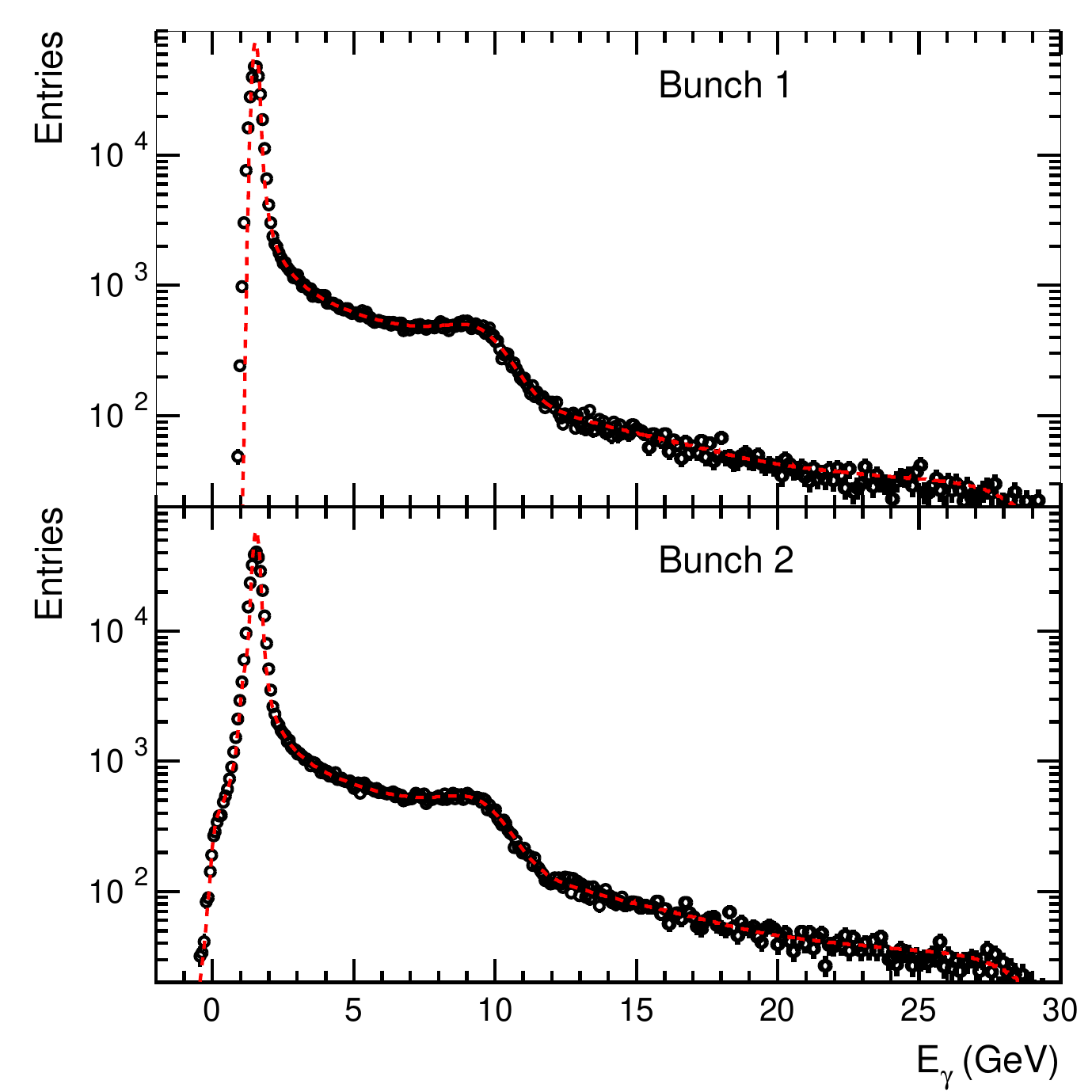}
\end{center}
\vspace{-3mm}
\caption{Difference in energy spectra between bunches 1 and 2 showing 
the pileup effect in bunch 2. The open dots show the data and the curves
the fit results.}
\label{fig:pileup}
\end{figure}

\item{\it Electronic sampling subtraction:}
Figure~\ref{fig:pileup} shows histograms of bunches 1 and 2 of the HERA beam: 
Bunch 2 shows a structure below the radiation peak.
This happens for full data acquisition runs representing 
about $40\%$ of the whole sample. 
It is due to a timing uncertainty at the start of data acquisition system: 
the signal is extracted from the maximum ADC sample of one bunch and 
the minimum ADC sample (baseline) from the previous bunch 
(see Sect.~\ref{seccalo}) instead of the subtraction of two ADC samples 
within one bunch. This results in a measured energy which is spoiled 
by the still active decay of
the preceding bunch signal after half an HERA clock 
($48\,{\rm ns}$). The measured energy is thus in this case
\begin{equation}\label{sampling}
E=E_{\rm bunch}-a_{\rm pileup}E_{{\rm bunch}-1}\,,
\end{equation}
where $a_{\rm pileup}$ is an attenuation parameter.
For the analysis, the bunches are treated chronologically and
the likelihood procedure always starts after empty bunches.
The energy spectrum of the first non-empty bunch (bunch 1) 
can thus be determined precisely without any correction. 
The energy spectrum of the next bunch (bunch 2) is corrected by a
convolution with the previous energy spectrum, according to
Eq.(\ref{sampling}), and so on for the following bunches.
After optimisation on a selected sample from the affected data we get 
a precise 
estimate of the attenuation parameter $a_{\rm pileup} = 0.057 \pm 0.002$.
This parameter remained constant within $\sim 10\%$ for all data taking. 
\end{description}

%\newpage
\section{Results of Polarisation Measurement} \label{sec:result}
All LPOL cavity data have been analysed with the polarisation results 
available to be compared with the corresponding measurements from the TPOL.
Here a few selected results are shown.

\begin{figure}[htbp]
\begin{center}
\includegraphics[width=.75\textwidth]{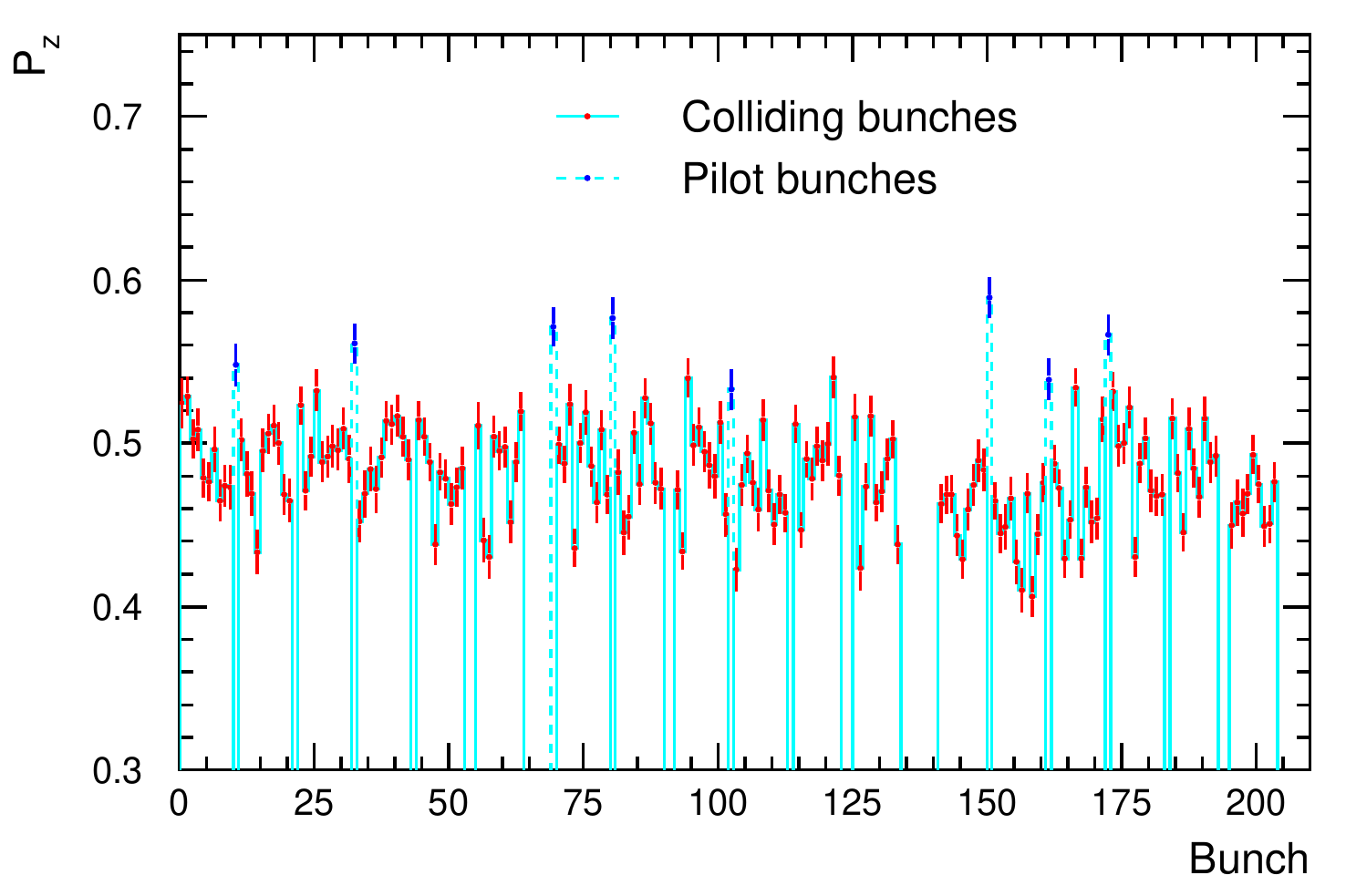}
\end{center}
\vspace{-3mm}
\caption{Bunch dependent polarisation measurement from the LPOL cavity 
averaged over 3 independent measurements corresponding to about one minute
duration. The error bars represent the statistical precision of the
measurement with the solid (dashed) histograms showing the colliding 
(pilot) bunch structure.}
\label{fig:bunch}
\end{figure}

The LPOL cavity provides a bunch dependent polarisation measurement every
$20\,{\rm s}$. The relative statistical precision is about $2\%$ per bunch 
per minute as shown in Fig.~\ref{fig:bunch} for a typical example.
In addition to the bunch structure, one also observes a significant 
bunch-to-bunch polarisation variation, not only between the colliding and 
non-colliding (pilot) bunches but also among the colliding bunches. 

\begin{figure}[htbp]
\begin{center}
\includegraphics[width=.8\textwidth]{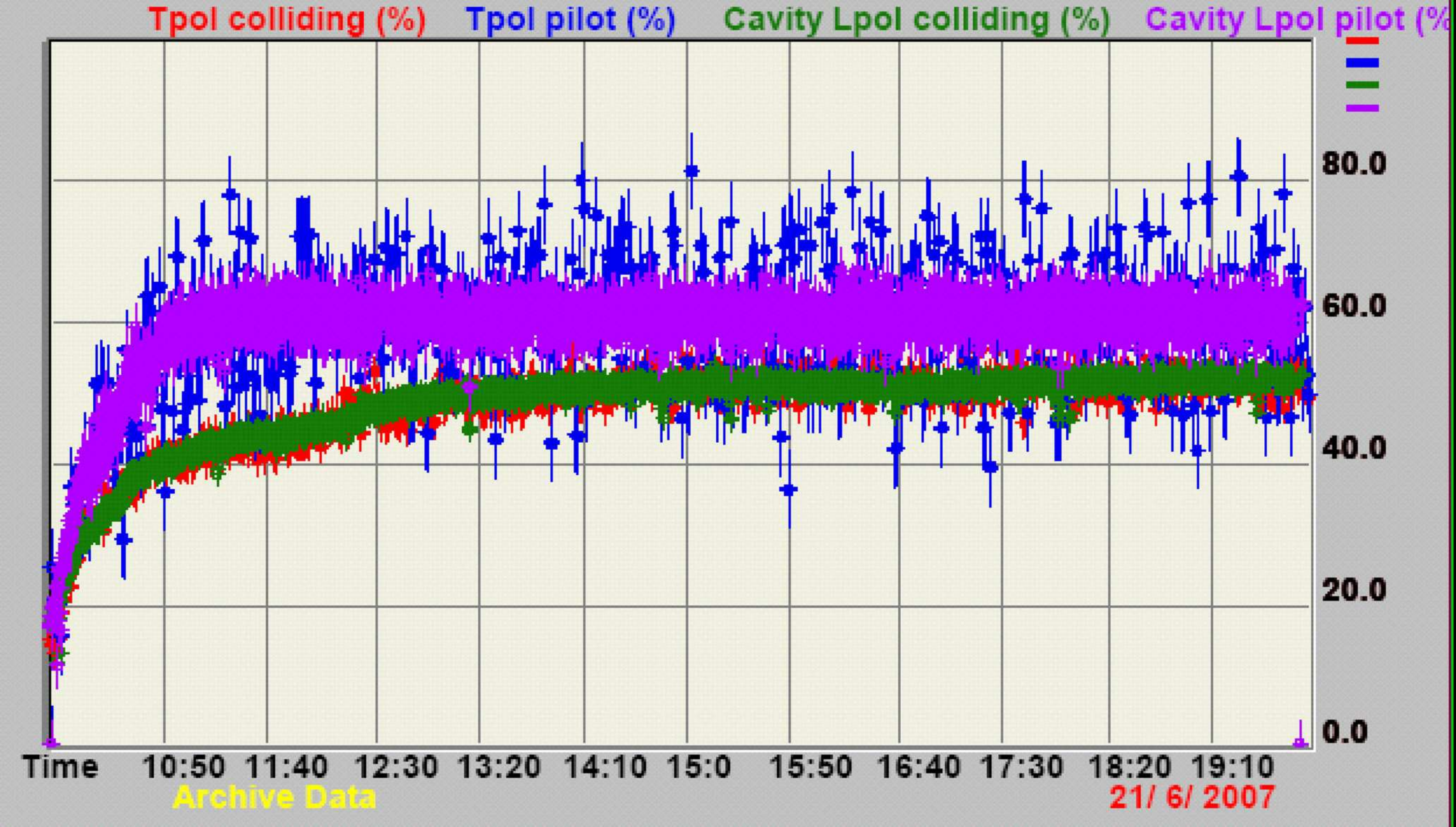}
\end{center}
\caption{Online polarisation measurements from the LPOL cavity (green 
and purple for the collidering and non-colliding (pilot) electron 
bunches, respectively) and the TPOL (red and blue).}
\label{fig:online}
\end{figure}

Figure~\ref{fig:online} shows an example of online polarisation measurements 
of one HERA luminosity fill provided by the LPOL cavity and TPOL polarimeters 
for both the colliding and pilot electron bunches. The online measurements
of the LPOL cavity were based on the same method as for offline measurements
described in previous sections.
The better statistical precision from the LPOL cavity polarimeter 
can be clearly appreciated.
Figure~\ref{fig:online} also shows that the polarisation values differ
significantly between the colliding bunches and pilot bunches.
Within the colliding bunches, the beam-beam effect is expected to vary 
depending on the proton bunch current, whereas
such effect is absent for the pilot bunches. These differences are indeed
observed from the bunch dependent polarisation measurement. 

\begin{figure}[htbp]
\begin{center}
\includegraphics[width=.65\textwidth]{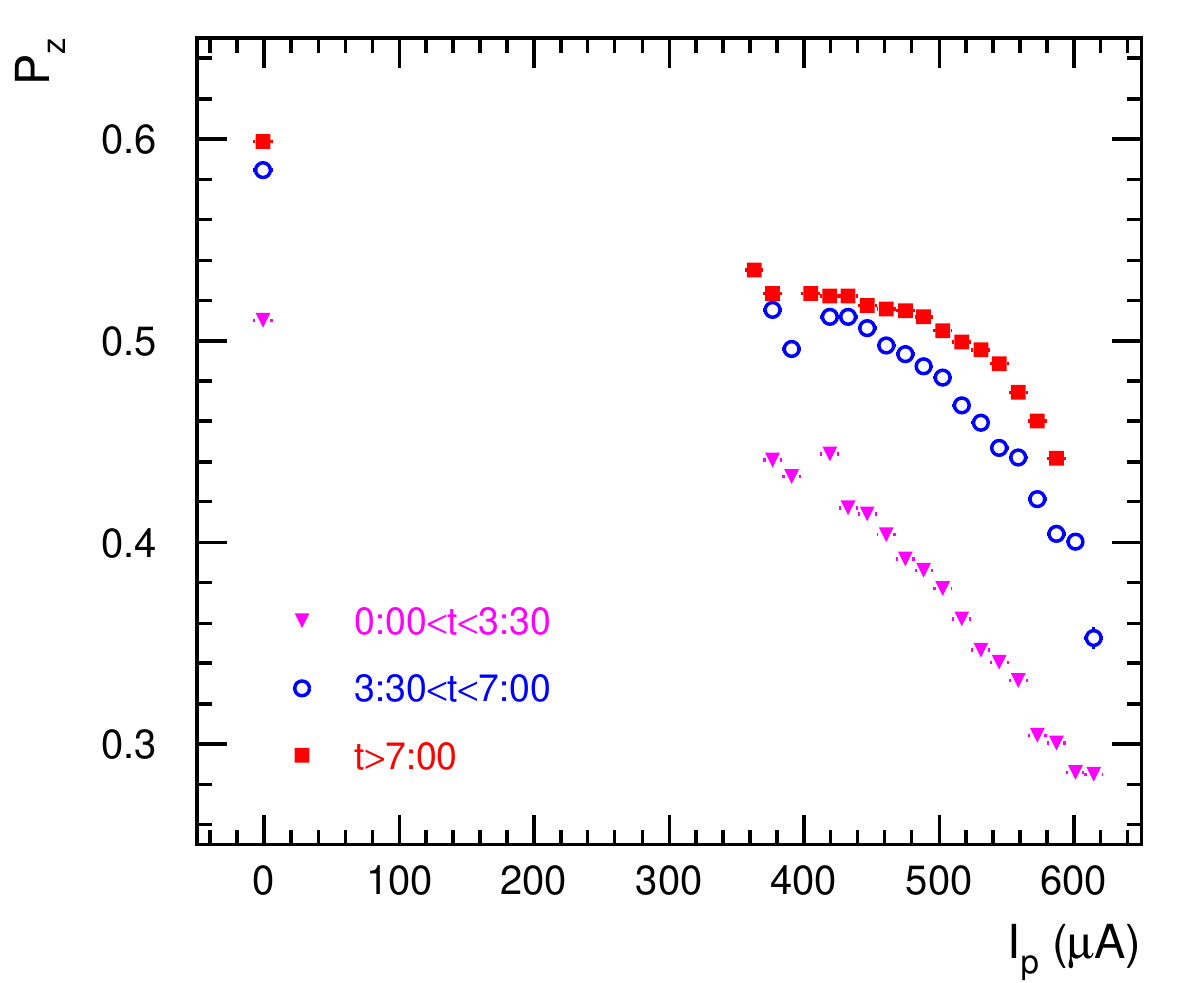}
\end{center}
\vspace{-3mm}
\caption{Measured polarisation dependence on the bunch-dependent proton
beam current and the time evolution.}
\label{fig:pcur}
\end{figure} 

Figure~\ref{fig:pcur} shows 
the dependence of the polarisation as a function of the proton beam current 
$I_p$ for a long HERA fill.
The data are split into three bins of the time since the start of the fill.
Here the colliding bunches have non-zero proton beam current $I_p\not =0$ 
and the pilot bunches are located at $I_p$=0.
Within the colliding bunches, one sees a dependence on $I_p$ which is strong
in the first half of the fill but becomes weaker in the second half. 
The bunch dependent polarisation measurements provided by the LPOL cavity 
allow the incorporation of all these effects into the relevant measurements
at HERA.

At the end of the HERA operation, some dedicated runs were taken
with the proton beam unfilled
and the electron beam undergoing a total of $13$ rise-time periods.
For each period, $P^\infty_z$, $\tau$ and $P^0_z$ (the polarisation at 
the beginning of the period) were extracted using
the averaged polarisation of all filled electron bunches with an estimated 
polarisation uncertainty $<0.2\%$.
The averaged polarisation $P_z(t)$ of each period was fitted using the formula
\begin{equation} 
P_z(t)=P^\infty_z +\left(P^0_z-P^\infty_z\right)e^{-\frac{t}{\tau}}\,.
\end{equation}

\begin{figure}[htbp]
\begin{center}
\includegraphics[width=.625\textwidth]{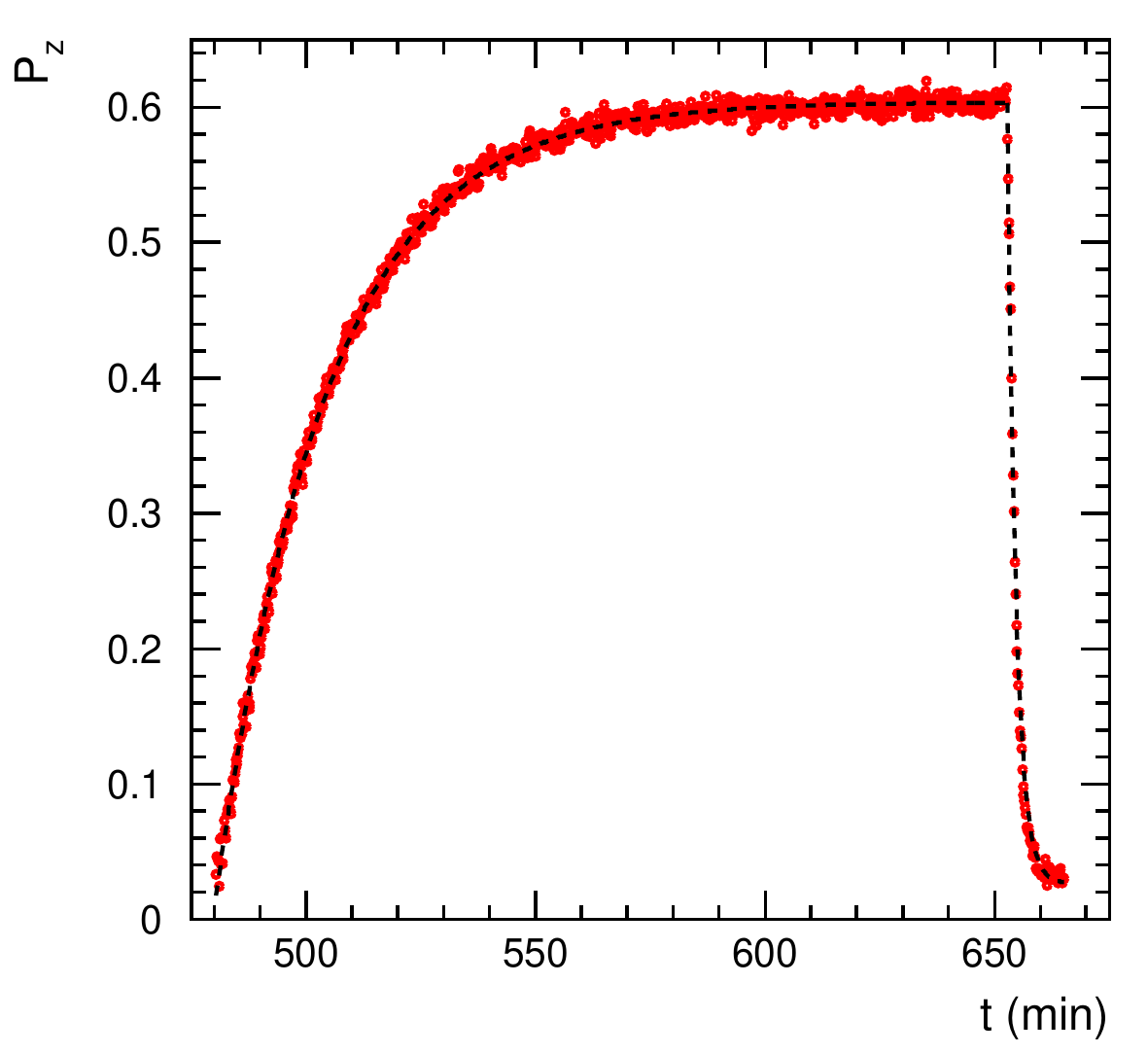}
\end{center}
\vspace{-3mm}
\caption{One example of the rise-time periods with the data points showing
the LPOL cavity measurements and the dashed curve the corresponding fit.}
\label{risefall}
\end{figure}

\begin{figure}[htb]
\begin{center}
\includegraphics[width=.625\textwidth]{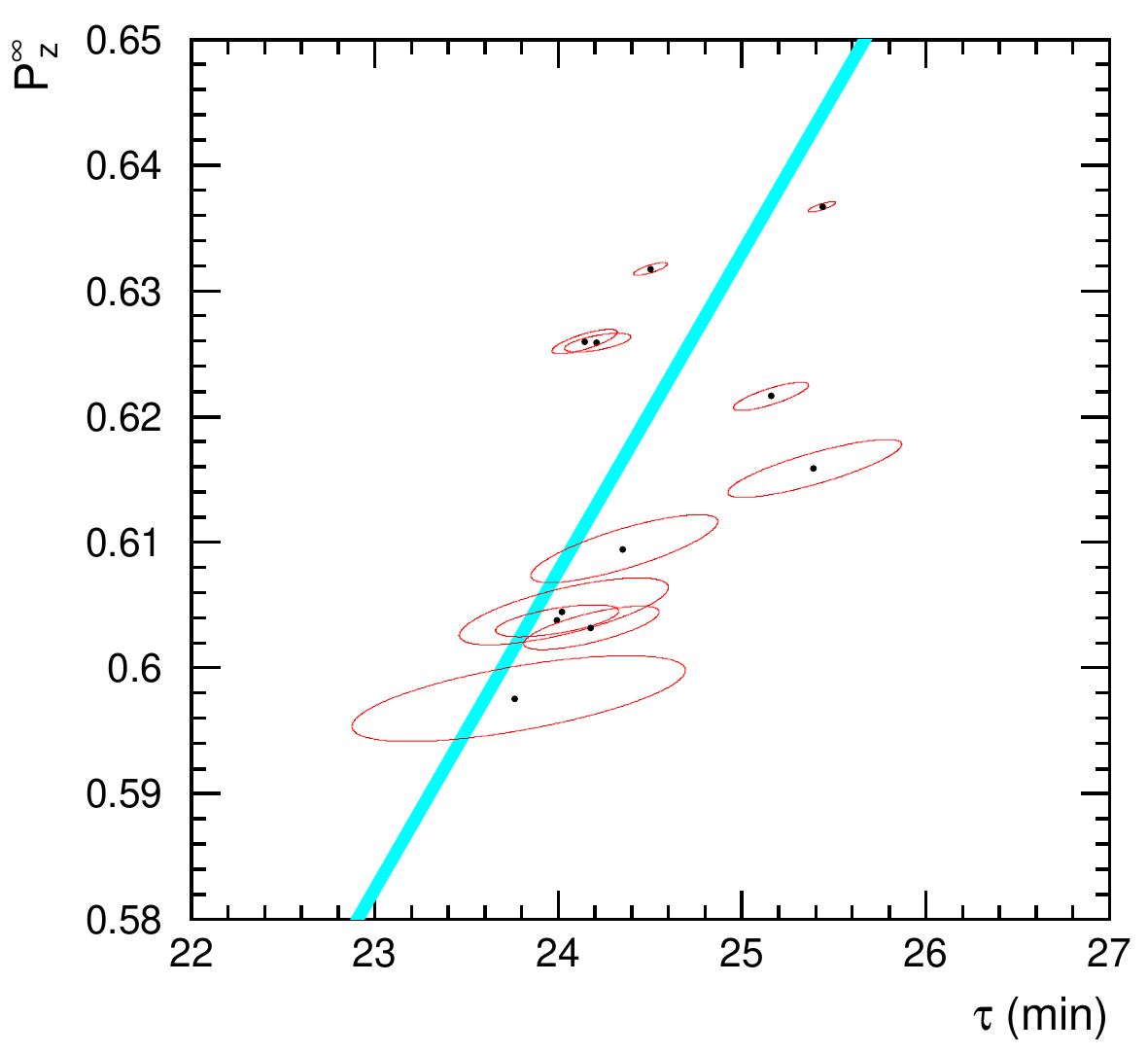}
\end{center}
\vspace{-3mm}
\caption{The fitted correlation between the maximum polarisation value
$P^\infty_z$ and the intrinsic rise-time $\tau$ for the $11$ long periods 
(ellipses) out of $13$ in total, in comparison with the expectation (straight 
line).}
\label{rise_S}
\end{figure}

One period is shown in Fig.~\ref{risefall} where the data points
correspond to the LPOL cavity measurements and the dashed curve the fit.
Towards the end, the beam was depolarised for starting a new rise-time period. 
Both the rise and rapid falling of polarisation were precisely measured.
Figure~\ref{rise_S} shows the plot of $P^\infty_z$ versus $\tau$ for the 11 
long rise-time periods. The straight line shows the ST prediction 
(Eq.(\ref{eq:st})), 
the result, $P^\infty_z/\tau = 0.02537  \pm  0.00004\,({\rm min}^{-1})$,
is about one standard deviation away from the expectation.

\newpage
\section{Systematic Studies}\label{sec:syst}
Detailed systematic studies have been performed using the full LPOL cavity 
data set, including dedicated data taking periods with non-standard setups.
Among different systematic sources,
we distinguish two types of uncertainties: one type (uncorrelated systematic
error) which is present for every single measurement but becomes negligible 
for a large integrated data sample since it behaves as a statistical 
uncertainty; the other type (correlated
systematic error) which is common for all the measurements.

\subsection{Uncorrelated Systematic Errors}\label{sec:uc_syst}
\noindent{\bf Systematics affecting independently each HERA fill (HERA beam
variations):}
The direction of the electron beam and the impact of the BCPs on 
the calorimeter affect
the calorimeter response leading to systematics. A devoted scan changing
the beam position and angle in the horizontal plane gives an error limit of 
$0.4\%$. However, to a first approximation this has to be divided by 
the square root of the number of fills (which represents about $100$
fills for the LPOL cavity data taking), thus becoming negligible.

\vspace{2mm}
\noindent{\bf Systematics related to detector parameters:}
Each data acquisition lasts for about $10$ seconds recording $220$ energy 
spectra for each bunch for a given laser polarisation, which is either 
left-handed or right-handed. The laser swapping frequency is matched
with that of the data acquisition system. Every 20 data acquisition
samples defines a microperiod. For each microperiod, a {\sc Minuit} fit is 
performed to define detector parameters using the measured spectra 
from those bunches which have at least three empty preceding bunches 
(thus the spectra are unaffected by the observed pileup feature mentioned 
in Sect.~\ref{sec:condition}).

To check the effect of the uncertainty of the fitted detector parameters, 
two independent methods are used:
\begin{itemize}
\item The parameters are varied according to the eigenvectors of the 
{\sc Minuit} Hessian matrix. The polarisations are refitted with the
new sets of parameters.
\item The polarisations are simply refitted with the set of parameters
determined from the next microperiod.
\end{itemize}
Both methods agree and give a relative uncertainty on the polarisation
measurement of about $0.5\%$. The corresponding systematic error is 
thus negligible due to more than 9000 microperiods of the LPOL cavity data 
taking.

\subsection{Correlated Systematic Errors}
\noindent{\bf The BGP and BBP cross-sections:}
They depend on the residual gas composition for BGP and on the beam pipe
temperature for BBP. Simulating a variation of these data has shown that
the induced systematics is negligible.

\vspace{2mm}
\noindent{\bf Calorimeter resolution and ADC to energy conversion parameterisations:}
The parameterisation
used for describing calorimeter resolution and ADC to energy conversion were
partially inspired from the ({\sc Geant}3) simulation. 
For the energy conversion, 
the possibility of leakage was taken into account. 
In fact, four other forms of parameterisation for the energy conversion were 
used and another Gaussian-like one was also tried for energy resolution. 
The 10 possibilities have been tested on a $1.5\%$ data subsample equally
distributed over the full sample and representing different situations
such as high and low BCP, BGP and SRP rates.
The solution retained was the one giving the best overall likelihood estimator,
{i.e.} the expression of Eqs.(\ref{wash}) and (\ref{E_ADC}).
The polarisation differences between the best solution and three other
good possibilities lead to a systematic error estimate of $0.4\%$.

\vspace{2mm}
\noindent{\bf Merging of the SRP peak:}
The likelihood estimator is very sensitive to the synchrotron radiation peak 
region (accounting for $\approx 80\%$ of the recorded entries 
per histogram)
which means that it is more affected by its fluctuations than by
the other interesting photons. Hence a merging of the energy histogram bins
in this region was made in order to reduce the effect while keeping 
the histogram number of entries constant.
By changing the way the bin merging
is made a systematic error of $0.4\%$ has been found on the used data subsample.

\vspace{2mm}
\noindent{\bf Left and right laser beam polarisations:}
The circular laser beam polarisation $S_3$ has been measured using
the optical ellipsometer with an uncertainty of $0.3\%$~\cite{mj}.
This has been however checked by a devoted scan measurement: in a period
of time short enough to expect that the electron polarisation
will only change slowly, five different MOCO position combinations
have been used to measure the polarisations. Each combination corresponds
to a different degree of circular laser polarisation.
The analysis of the scan data leads to a systematics error consistent with 
the above value.

\vspace{2mm}
\noindent{\bf Electronic sampling subtraction:}
The same attenuation factor $a_{\rm pileup}$ has been used for all the data 
samples. In fact it has been measured only on a dedicated subsample. 
Its size and stability
showed no necessity to have another strategy. 
For the whole sample the policy was to measure the polarisation imposing 
$a_{\rm pileup}=0$
and calculate the correction by use of the dedicated subsample.
In fact some dependence of the correction with respect to the BGP and
BCP yields was found and used.
Taking into account the width of $a_{\rm pileup}$, 
the estimated error on the correction
and the relative number of affected measurements, the corresponding systematic
error has been estimated to be $0.4\%$.

\vspace{2mm}
\noindent{\bf Calorimeter position scan:}
In the nominal data taking, the calorimeter is centered 
around the scattered photon beam line. During a dedicated run, 
the calorimeter has been moved both vertically and 
horizontally. The polarisation
measurement is found to be stable within $0.4\%$ for both directions.

\subsection{Overall Systematic Uncertainty}
The systematic uncertainties of all considered error sources are
summarised in (Table~\ref{tab:syst}). It should be pointed out however that
\begin{table}[htb]
\caption[.]{\label{tab:syst}
  Estimation of the relative systematic errors of the LPOL cavity
  polarimeter for an individual polarisation measurement.}
\begin{center}
\begin{tabular}{lc}
\hline
Source & $\Delta P/P (\%)$\\ \hline
\multicolumn{2}{c}{Uncorrelated errors}\\\hline
HERA beam variations & $0.4$ \\
Detector parameters  & $0.5$ \\\hline
\multicolumn{2}{c}{Correlated errors}\\\hline
BGP and BBP cross-sections & negligible \\
Calorimeter resolution and ADC to energy conversion & $0.4$ \\
Merging of the SRP peak & $0.4$ \\
Laser polarisation circularity & $0.3$ \\
Electronic sampling subtraction & $0.4$ \\
Calorimeter position scan (horizontal) & $0.4$ \\
Calorimeter position scan (vertical) & $0.4$ \\ \hline
%Total & 0.9 \\ \hline
\end{tabular}
\end{center}
\end{table}
\begin{itemize}
\item
The scan analysis method assumes the polarisation to vary linearly
with time during the scan and the scan is performed in a relatively 
short period of time.
This induces an intrinsic investigation limit of $\sim 0.5\%$
so the real value of the error could (and sometime is expected to) be
much smaller.
\item
The uncorrelated errors should be added in quadrature but here some are 
correlated in a somewhat uncertain way. This renders the summing process
also rather uncertain.
\end{itemize}
Bearing all that in mind an overall relative systematic error of $0.9\%$ 
is quoted on the LPOL cavity polarisation measurement for a large integrated
period. For an individual measurement in a short time period, 
the corresponding uncertainty is $1.1\%$.

%%%%%%%%%%%%%%%%%%%%%%%%%%%%%%%%%%%%%%%%%%%%%%%%%%%%%%%%%%%%%%%
\section{Summary}\label{sec:summary}
A new polarimeter for precisely measuring the longitudinal polarisation
of the electron beam at HERA has been constructed and successfully
operated. The polarimeter employs a Fabry-Perot cavity for enhancing
the laser intensity by more than three orders of magnitude so that
the electron-photon interaction rate is sufficiently large to reach 
the few photon mode.
A fast data acquisition system has also been developed to record all 
back-scattered photons produced every $96\,{\rm ns}$ for every electron
bunch.

The measurement of the electron beam polarisation is reported for
the first time in the few photon mode leading to a statistical precision
of $2\%$ per bunch per minute. This offers an improvement 
over the other two HERA polarimeters 
which are limited by either lower laser intensity or smaller electron-photon 
interaction rate.

Detailed systematic studies have been performed resulting in a
total relative systematic uncertainty of about $1\%$, which is 
a factor of $2-3$ smaller than the precision quoted currently by the
other polarimeters at HERA. To reach such a small systematic uncertainty,
we have used the possibility to describe
the few photon energy spectra from first principles by convoluting 
the signal and background QED processes with the detector effects. 
Our major observation is that the detector parameters, 
used to relate the theoretical energy spectra
to the measured ones, could be determined once every $3$ minutes of data taking
independently of the electron beam polarisation measurement. 
This allowed us to account for detector response variations as functions of 
the photon entrance position in the calorimeter and aging effects.

These precise polarisation measurements can be used as a reference
to cross-calibrate the other polarimeters and to help resolve
some of the remaining discrepancies observed between 
the two other HERA polarimeters.

\section*{Acknowledgment}
We are grateful to all members of the POL2000 group for their support
and discussions. The support from the HERA machine group, and
the H1, HERMES and ZEUS experiments is also warmly acknowledged.
We thank A.~Airapetian, M.~Ait-Mohand, C.~Cavata, I.~Cheviakov, R.~Fabbri, 
M.~Hensel, Y.~Holler, M.~Klein, J.~Ludwig and
C.~Vall\'ee for their help at different stage of the project.

\end{document}